\documentclass[aps, twocolumn,groupedaddress,amsmath,ams-fonts,showpacs,dvipdfmx]{revtex4}

\usepackage{graphicx}
\usepackage{dcolumn}
\usepackage{bm}
\usepackage{here}
\usepackage{color}
\usepackage{relsize}

\begin{document}

\title{Giant Nernst and Hall Effects in Chiral Superconductors due to Berry Phase Fluctuations}

\author{Hiroaki Sumiyoshi$^1$}
\author{Satoshi Fujimoto$^{1,2}$}%
\affiliation{$^1$Department of Physics, Kyoto University, Kyoto 606-8502, Japan}
\affiliation{$^2$Department of Materials Engineering Science, Osaka University, Toyonaka 560-8531, Japan}

\date{\today}

\begin{abstract}

We consider the Nernst and Hall effects in fluctuation regime of chiral superconductors above transition temperatures,
that are raised not by conventional Lorentz force, but by asymmetric scattering due to fluctuations of  the Berry phase of the Bogoliubov-de Gennes Hamiltonian.
It is found that these effects can be more significant than conventional ones for cleaner samples, exhibiting qualitatively distinct behaviors.
The results provide systematic and comprehensive understanding 
for recent experimental observations of the Nernst effect in a clean URu$_2$Si$_2$ sample, which is suggested to be a chiral superconductor.

\end{abstract}

\maketitle

In a certain class of superconductors, fluctuations toward ordered states above transition temperatures give rise to dramatic effects
on many-body electron states.
It is known that a powerful probe for such phenomena is the Nernst effect.
For instance, giant Nernst signals have been observed in near and above transition temperatures $T_c$ of
cuprate high-$T_c$ superconductors \cite{Xu} and dirty superconducting thin films \cite{Pourret}.
In normal metals, the Nernst signal is generally weak owing to the Sondheimer cancelation \cite{Behnia}
and then, these unexpected experimental observations inspired succeeding extensive studies, leading to various theoretical proposals such as
scenarios based on
short-lived Cooper pairs \cite{Ussishkin2002},
Josephson electromotive force due to the vortex motion \cite{WangPRB2001},
and 
strong coupling with antiferromagnetic fluctuations \cite{Kontani}.

In this letter, we propose an unconventional mechanism for the giant Nernst effect above $T_c$ in {\it chiral superconductors}, which
has not been discussed so far.
In chiral superconductors, time-reversal symmetry (TRS) is spontaneously broken, and total angular momentum carried by Cooper pairs is nonzero.
The ``chirality'' of this superconducting state
is characterized by the Berry phase of the Bogoliubov-de Gennes mean-field Hamiltonian, which is
an Aharonov-Bohm (AB) phase whose adiabatic parameters are the wave number \cite{VolovikHe}.
In the superconducting phase below $T_c$, the intrinsic magnetic field induced by it causes
exotic transverse transport phenomena under zero external magnetic field,
such as the Kerr effect \cite{Furusaki2001, RoyKallin}, which was observed in Sr$_2$RuO$_4$ \cite{Xia}, 
and the anomalous thermal Hall effect, which was theoretically predicted \cite{RG, Nomura, Sumiyoshi}.
It is natural to expect that also in the superconducting fluctuation regime above $T_c$, characteristic transverse transport phenomena can be induced by fluctuations of the chirality or the Berry phase.
We investigate this possibility, and clarify a novel mechanism of the giant Nernst and Hall effects above and near $T_c$, caused by Berry phase fluctuations.
In this scenario, quasiparticles are scattered asymmetrically by fluctuating Cooper pairs with angular momentum, even without Lorentz force,  
and then, such effects can be regarded as an analog of the skew scattering process of the anomalous Hall effect,
which is caused by a spin-orbit coupling involving impurity scattering \cite{NN_AHE},
but a major difference is that the scattering kernels are dynamical in this case.

There are several candidate systems for chiral superconductors, e.g. 
Sr$_2$RuO$_4$, URu$_2$Si$_2$, doped graphene/silicene, SrPtAs, and Na$_x$CoO$_2\cdot y$H$_2$O
\cite{Mackenzie, KasaharaPRL, KasaharaNJP, Yano, Nandkishore, Liu, Biswas, Fischer, Kiesel}.
Among them, the heavy-electron superconductor URu$_2$Si$_2$, whose pairing symmetry is suggested to be chiral $d_{zx} \pm {\rm i} d_{zy}$ \cite{KasaharaPRL, KasaharaNJP, Yano},
is one of the most promising one for the realization of the above-mentioned mechanism, 
because, for this system,  
strong superconducting fluctuation effects have been experimentally observed, which may be attributed to small energy scale
raised by heavy effective mass and the reconstruction of electronic structures in the so-called hidden order phase
\cite{Okazaki}. 
Thus, in this letter, we mainly focus on this system, though
our theory is also applicable to other chiral superconductors with minor modifications.

Our approach is based on microscopic model calculations utilizing linear response theory for the Nernst and  Hall effects.
The Hall conductivity is given by the Kubo formula:
$\sigma_{\alpha \beta}=\left. S^{R}_{\alpha \beta}(\omega)/(-{\rm i} \omega) \right|_{\omega \to 0}$,
where $S^{R}_{\alpha \beta}(\omega)$ is the retarded current-current correlation function.
On the other hand, the case of the Nernst effect is more involved. As pointed out by previous studies,
one needs to take account of contributions from magnetization $\bm{M}$ in addition to
those from the Kubo formula \cite{Luttinger, Obra, Smrcka, Cooper}.
Then, the Nernst conductivity (Peltier coefficient) is,
\begin{eqnarray}
 \alpha_{\alpha \beta} &=& \alpha_{\alpha \beta}^{Kubo} + \alpha^{mag}_{\alpha \beta} , \nonumber \\
 \alpha^{mag}_{\alpha \beta} &=&  \epsilon_{\alpha \beta \gamma} M^{\gamma}/T .
 \label{alphaxy}
\end{eqnarray}
Here, $\alpha_{\alpha \beta}^{Kubo} $ is the Kubo term given by the heat-current-charge-current correlation function.
We apply diagram techniques to calculate these transport coefficients.

{\it Model} ---
The Hamiltonian with which we start is an effective model for the superconducting state of URu$_2$Si$_2$ which belongs to 
 $E_g$ representation of  the point group $D_{4h}$ \cite{SigristUeda}:
\begin{eqnarray}
&&H=\sum_{\bm{ k} \sigma} \xi_{ \bm{k}}  c^{\dag}_{ \bm{k}\sigma}  c_{ \bm{k}\sigma}  \nonumber\\
&&\,  - g \sum_{ \bm{k,k',q}} V(\bm{ k,k'}) c^{\dag}_{ \bm{k+q}/2\uparrow} c^{\dag}_{-\bm{ k+q}/2\downarrow}  c_{-\bm{ k'+q}/2\downarrow} c_{\bm{ k'+q}/2\uparrow} , \nonumber \\
\end{eqnarray}
where, for simplicity, we take a spherical Fermi surface, $\xi_{\bm{ k}} = \bm{ k}^2/2m -\mu $, and 
the effective attractive interaction is given by $V(\bm{ k},{\bm k'})=15(k_z k_x k'_z k'_x + k_z k_y k'_z k'_y )/k^4_{F}$.
It is the model for
 the chiral $d_{zx} \pm {\rm i} d_{zy}$ superconductor URu$_2$Si$_2$.
In the chiral superconducting phase, TRS is spontaneously broken and the gap function takes the form $\Delta(\bm{ k}) \propto k_z (k_x + {\rm i} k_y)$ (or $k_z (k_x - {\rm i} k_y)$),
which is caused by an effective attractive interaction, $V^{+}(\bm{ k},{\bm k'})=\phi(\bm{ k})\phi^{\dag}(\bm{ k'})$ (or $V^{-}(\bm{ k},{\bm k'})=\phi^{\dag}(\bm{ k})\phi(\bm{ k'})$),
where the pairing symmetry function reads $\phi(\bm{ k})= \sqrt{15/2} k_z (k_x + {\rm i} k_y)/k^2_{F}$.
Note that $V(\bm{ k},\bm{ k}') =V^{+}(\bm{ k},{\bm k'})+V^{-}(\bm{ k},{\bm k'})$.
The channel $V^{+(-)}$ is associated with the chirality $C=+1$ ($-1$), and each channel breaks TRS.
However, we concentrate on
transport phenomena above $T_c$, in fluctuation regime, where two channels are degenerate,
and therefore TRS is {\it not} spontaneously broken.

{\it Nernst and Hall Effects} ---
Generally, to induce transverse transport phenomena such as the Nernst and Hall effects,
it is necessary to break TRS.
In fluctuation regime above $T_c$, TRS is not spontaneously broken, 
and then a magnetic field is necessary
unlike spontaneous Kerr and thermal Hall effects in the chiral superconducting phases \cite{Furusaki2001, RoyKallin, Xia, RG, Nomura, Sumiyoshi}.
Due to a magnetic field, the Lorentz force on quasiparticles and fluctuating Cooper pairs is generated
and causes conventional transverse transport phenomena \cite{Larkin}.
In addition, in the case of chiral superconductors, the magnetic field also causes ``polarization" of chirality due to a magnetic filed-chirality (MC) coupling;
i.e. the difference in the weights of two superconducting fluctuation channels is induced.
The chirality-polarized superconducting fluctuations give rise to asymmetric scattering of electrons resulting in the anomalous Nernst and Hall effects (ANE and AHE) without Lorentz force, which are the main subjects of this letter (See Eqs.(\ref{final_Nernst_Kubo}), (\ref{final_Hall}), and (\ref{mag_current}) below, which constitute the main results).


First, we discuss the chirality polarization by evaluating the superconducting fluctuation propagator.
Under a uniform magnetic field  $\bm{ H} = (0, 0, H)$, the fluctuation propagators of chiral $d_{zx}\pm d_{zy}$-channels (correspond to $C = \pm 1$, respectively) is given by  \cite{sup}:
\begin{eqnarray}
&&\tilde{L}^{-1}_C({\bm x},{\bm y},\omega_q;H)=-\frac{\delta({\bm x}-{\bm y})}{g} + \tilde{\Pi}_C({\bm x},{\bm y},\omega_q;H), \nonumber \\
&&\tilde{\Pi}_{C} (\bm{ x,y}, \omega_q; H ) = e^{-{\rm i}2 e \Phi(\bm{ x}, \bm{ y})  } \nonumber \\
&& \times \left[  \Pi (\bm{ x-y}, \omega_q ; H) - C  \frac{5eH}{4k^2_{F}} \Pi' (\bm{ x-y}, \omega_q ;H) \right] , 
 \label{pp}
\end{eqnarray}
where $\Pi$ and $\Pi'$ are ``core" bare particle-particle susceptibilities (BPSs) of $d_{zx} \pm {\rm i} d_{zy}$- and $p_z$-wave channels, respectively,
which preserve translation, gauge, and $c$-axis rotation invariances, 
$\omega_q$ is the bosonic Matsubara frequency,
and the AB-phase, $\Phi(\bm{ x}, \bm{ y})=   \int_{\bm{x}}^{\bm{y}}  \bm{ A (r)} d\bm{ r}$, is defined as an integral of the vector potential along a straight line.
Here we used the fact that the one-particle Green function in a magnetic field is given by
$\tilde{G}(\bm{ x}, \bm{ y}, \varepsilon_n ; H)=e^{-{\rm i}e \Phi(\bm{ x}, \bm{ y})  }G_{core}(\bm{ x}- \bm{ y}, \varepsilon_n;H)$,
where $G_{core}$ is the core Green function, which has translation, gauge, and $c$-axis rotation invariances \cite{Khodas2003, KarenHall}.
We note that this expression (\ref{pp}) is applicable to arbitrary magnitude of magnetic fields and for any gauge conditions.
The remarkable point is the existence of the chirality-dependent term, $-C(5eH/4k^2_{F}) \Pi' $,
which changes the amplitude of the BPS, reflecting the polarization mechanism due to the MC-coupling.
The MC-coupling raises (lowers) the transition temperature of the $C=-1$ ($+1$) state, which has orbital magnetic moment parallel (antiparallel) to the $c$-axis,
in contrast to the AB-phase, which reflects the orbital depairing effect, and always lowers the transition temperature \cite{HW}.
Moreover, the MC-coupling induces paramagnetism discussed later.





\begin{figure}
 \begin{center}
  \includegraphics[width=80mm]{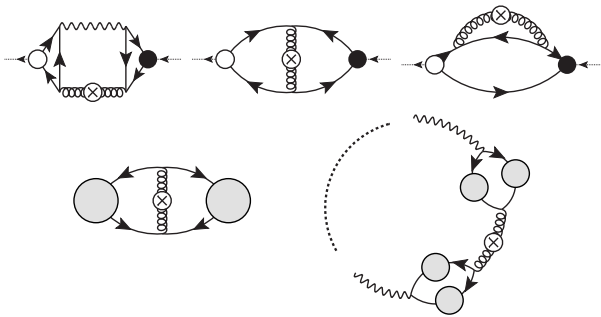}
 \end{center}
\caption{{\it Upper panel}:  
AL, MT, and DOS diagrams. 
The AL and DOS diagrams have the mirror image counterparts.
Wavy lines and curly lines with crossed circles represent the fluctuation propagator in zero magnetic field, $L$, and the chirality-polarized one, $\tilde{L}'_C$, respectively,
where their definitions are given in \cite{sup}.
Solid lines with arrows are the one-particle Green functions. 
Open circles represents electric current vertex, 
and bullets represent energy current vertex (electric current vertex), 
for $\alpha_{xy}$ ($\sigma_{xy}$).
{\it Lower panel}: Diagrams in which the information of the chirality disappears, resulting in vanishing contributions to the Nernst or Hall effects.
Shaded circles represent any diagrams without fluctuation propagators and the two current vertices are inserted into any propagators.}                                                                                                                                                                                                                                         
 \label{AL_MT_DOS}
\end{figure}

\begin{figure}
 \begin{center}
  \includegraphics[width=70mm]{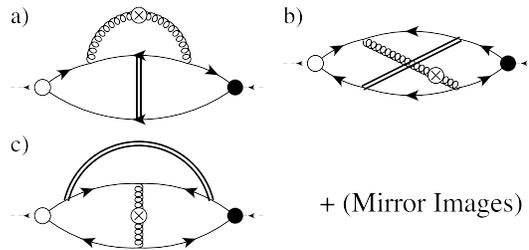}
 \end{center}
\caption{Diagrams which contribute to the ANE and AHE raised by Berry-phase fluctuation mechanism. The double lines represent the renormalized four-point vertex, $W(\bm{ k}, \omega_k)$ due to electron-electron interaction.}                                                                                                                                                                                                                                      
 \label{news}
\end{figure}

Using the fluctuation propagator, Eq. (\ref{pp}), we calculate 
the Nernst and Hall conductivities.
Note that up to the linear order in $H$, 
we can systematically separate whole contributions 
into two parts: one corresponding to the conventional contribution due to Lorentz force on quasiparticles and fluctuating Cooper pairs,
and the other one associated with the ANE and AHE caused by the Berry-phase fluctuation mechanism.
As will be shown below, the latter contribution dominates over the former one for clean samples.
Thus,  we focus on the latter in the following.
The detail of the calculation based on diagrammatic techniques is presented in the Supplemental Material \cite{sup}.
We sketch briefly a basic idea of the derivation for the Nernst and Hall conductivities.
It is found that the three diagrams which give leading-order contributions in conventional theories, i.e.
the Aslamazov-Larkin (AL), Maki-Thompson (MT), and density-of-states (DOS) diagrams (upper panel in Fig. \ref{AL_MT_DOS}) \cite{Larkin},
do not contribute in this case,
and generally, all contributions from diagrams belonging to the classes of the lower panel in Fig. \ref{AL_MT_DOS} are zero \cite{sup}. 
The lowest order diagrams which do not belong to these classes and give nonzero contributions are depicted in Fig. \ref{news}.
In these diagrams, inelastic scattering processes due to electron-electron interaction represented by a renormalized four-point vertex, $W(\bm{ k}, \omega_k)$ (double line), 
are included.
To carry out calculations explicitly, we postulate a simple model: $W(\bm{ k}, \omega_k)=W_0 /(1+|\omega_{k}|/\Gamma )$, i.e. an interaction mediated via
a short-range antiferromagnetic spin fluctuation,
where $W_0$ is a constant
and $\Gamma$ is the energy scale of spin fluctuations.
In fact, for URu$_2$Si$_2$, 
a short-range antiferromagnetic spin fluctuation 
exists  in the hidden order phase as clarified by inelastic neutron scattering measurements \cite{Bourdarot, Wiebe}.
Thus, the above assumption for $W(\bm{ k}, \omega_k)$ is legitimate.
However, we stress that our final results are qualitatively not changed by specific form of $W(\bm{ k}, \omega_k)$,
as will be discussed later.


Then, we obtain the Kubo terms of the Nernst and Hall conductivities in clean limit, near $T_c$, and in the linear order of $H$ \cite{sup}:
\begin{eqnarray}
\frac{\alpha^{Kubo}_{xy \, chiral}}{H} &=& \frac{f\left( \frac{2\pi T}{\Gamma} \right)}{2304}
\frac{ \tau^2 e^2 W_0 v^4_F \Lambda }{\xi^4 g k_F^2 T^2} \left(1-\frac{3 \pi }{4} \frac{ \sqrt{ \varepsilon}}{\xi\Lambda} \right),  \label{final_Nernst_Kubo} \\
\frac{\sigma_{xy \, chiral}}{H} &=& \frac{5 f\left( \frac{2\pi T}{\Gamma} \right)}{1152}\frac{\tau^2 e^3  W_0 v_F^3  \Lambda }{\xi^4 g k^3_F T}  \left(1-\frac{3 \pi }{4} \frac{ \sqrt{ \varepsilon}}{\xi\Lambda} \right). \label{final_Hall}
\end{eqnarray}
Here, $\varepsilon = \log T/T_c $, $v_F$ is the Fermi velocity,
$\xi = \sqrt{- \psi''(1/2)/6}(v_F/4\pi T) $ is the  coherence length,
$\psi$ is the digamma function,
$\tau$ is the electron scattering time due to impurities and electron-electron scattering,
$\Lambda$ is the cutoff of the momentum of superconducting fluctuation propagator, which is the same order as $1/ \xi$,
and $f(2\pi T/\Gamma)$ is a dimensionless function, whose definition and numerical estimations are given in \cite{sup}.

Now, we discuss the magnetization contribution in Eq.(\ref{alphaxy}).
The magnetization due to chirality-polarized superconducting fluctuations is of interest not only because of its contribution to the Nernst effect, but also because of its unique magnetic property; i.e. the polarization of chiral superconducting fluctuation channels causes paramagnetism in contrast to diamagnetism due to fluctuating Meissner currents observed
in general superconductors \cite{Larkin}.
The calculation is performed with the free energy of chiral superconductors above $T_{c}$:
\begin{eqnarray}
F[H] = T \sum_{\omega_q, C= \pm 1} {\rm Tr} {\rm ln} (-\hat{\tilde{L}}_{C}^{-1}(\omega_q; H)),
\end{eqnarray}
where $\hat{\tilde{L}}_{C}^{-1}(\omega_q; H)$ is the matrix whose indices are spatial coordinates, ${\bm x}$ and ${\bm y}$,
and matrix elements are $\tilde{L}_{C}^{-1}({\bm x}, {\bm y}, \omega_q; H)$, Eq. (\ref{pp}).
From this free energy, we obtain the magnetic susceptibility $\chi = \chi_{dia} + \chi_{chiral}$ with
\begin{eqnarray}
&& \chi_{chiral} = \frac{25 e^2 T}{64 \pi k_F^4 \xi^3 (N(0) g)^2 \varepsilon^{1/2}} >0,
\end{eqnarray}
where $\chi_{dia}$ is the diamagnetic term due to fluctuating Meissner currents observed in general superconductors \cite{Galitski2001, Larkin},
and $\chi_{chiral}$ is the paramagnetism term mentioned above \cite{sup}.
Then, the magnetization current contribution inherent in chiral superconductors is
\begin{eqnarray}
\frac{\alpha^{mag}_{xy\, chiral}}{H}&=& \frac{\chi_{chiral}}{T}= \frac{25 e^2}{64 \pi k^4_F \xi^3 (N(0) g)^2 \varepsilon ^{1/2}} . \label{mag_current}
\end{eqnarray}

The total Nernst conductivity due to Berry-phase fluctuations is given by sum of Eqs. (\ref{final_Nernst_Kubo}) and (\ref{mag_current}),
which constitute our main results.

\begin{figure}
 \begin{center}
  \includegraphics[width=75mm]{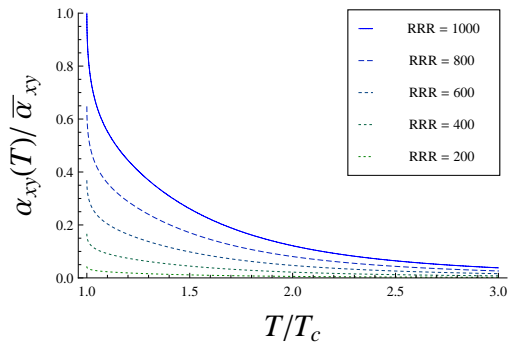}
 \end{center}
\caption{(Color online). $\alpha_{xy}$ raised by the Berry-phase fluctuation mechanism versus $T/T_c$
for several values of RRR.
The magnitudes of $\alpha_{xy}$ are normalized by the value of the most clean one at $T_c$, $\bar{\alpha}_{xy} := \alpha_{xy}(T_c; {\rm RRR} = 1000)$.
We used the material parameters of URu$_2$Si$_2$ \cite{comment_on_alpha}.}                                                                                                                                                                                                               
 \label{fig:alpha}
\end{figure}

{\it Discussions} ---
We now discuss several important features of Eqs. (\ref{final_Nernst_Kubo}) and (\ref{mag_current}).
The critical behavior of the magnetization current contribution, $(\ref{mag_current})$, given by $\propto (T-T_c)^{-1/2}$, 
is the same as
that of the AL term of the Nernst conductivity, which is also obtained from a time-dependent Ginzburg-Landau (TDGL) equation \cite{Ussishkin2002, Ussishkin2003}.
On the other hand, the critical behavior of the Kubo contribution, $(\ref{final_Nernst_Kubo})$, 
is less singular, $\propto (const. - \sqrt{T-T_c})$.
However, we note that the dependence on scattering time $\tau$ of Eq. $(\ref{final_Nernst_Kubo})$, which is proportional to $\tau^2$, 
is quite distinct from
any conventional fluctuation-induced corrections to the Nernst coefficient previously studied so far.
For instance, there is no $\tau$-dependence in the contribution to $\alpha_{xy}$ that obtained by dynamics of boson fields (i.e. fields of Cooper pairs), 
such as the scenarios of short-lived Cooper pairs (i.e. the conventional AL term) \cite{Ussishkin2002} and the vortex motion \cite{WangPRB2001}.
This is simply because that dynamics of bosons do not involve quasiparticle scattering time. 
Also, it is known that contributions from electron dynamics influenced by the fluctuation boson field,
including the MT and DOS terms, do not yield $\tau$-dependent $\alpha_{xy}$
\cite{Serbyn2009, MichaeliEPL}.  
Thus, for sufficiently clean samples with large $\tau$, 
the Kubo term $\alpha_{xy\, chiral}^{Kubo}$ of the Berry-phase fluctuation mechanism significantly dominates over
contributions from the AL, MT, and DOS terms of the Nernst conductivity raised by conventional Lorentz force.
Furthermore, because of the $\tau$-dependence, the Kubo term $\alpha_{xy \, chiral }^{Kubo}$ is also much more enhanced than the magnetization term (\ref{mag_current}) for cleaner samples.
Thus, the leading term of the Nernst conductivity for clean chiral superconductors is given by $\alpha^{Kubo}_{xy \,chiral}$.
The unusual $\tau$-dependence of $\alpha^{Kubo }_{xy \, chiral}$ combined with an increasing behavior  
for $T$ approaching to $T_c$,
as shown in Eq. (\ref{final_Nernst_Kubo})
characterizes the distinct feature of the Berry-phase fluctuation mechanism.
In Fig. \ref{fig:alpha}, we plot typical temperature dependences of Eq. (\ref{final_Nernst_Kubo}) for several values of $\tau$ 
parametrizing the residual resistivity ratio (RRR) of samples.
Here, we used material parameters of URu$_2$Si$_2$
\cite{comment_on_alpha}, and the calculation was achieved by using an approximation scheme explained in \cite{sup}.
As discussed above, $\alpha_{xy}$ exhibits remarkably strong enhancement in the vicinity of $T_c$
for cleaner systems. 
It is an intriguing feature issue to test our theory for real materials.
On the other hand, the Hall conductivity, 
Eq. (\ref{final_Hall}),
has the same characteristic $\tau$-dependence, $\propto \tau^2$, as $\alpha_{xy  \, chiral}^{Kubo}$,
and, moreover, is nonzero even when the electronic band is particle-hole symmetric.
This point is quite different from conventional contributions derived from TDGL equation or, equivalently, the AL term,
which requires particle-hole band asymmetry: $\partial T_c/ \partial \mu \neq 0$ (equivalently $\partial N(0)/ \partial \mu \neq 0$ or $\partial g/ \partial \mu \neq 0$) \cite{AronovHall}.
However, it would be rather more difficult to detect the Hall effect than the Nernst effect,
because normal Hall currents of conventional Fermi-liquid quasiparticles dominate for charge transport.

These $\tau$-dependences of the Peltier and Hall coefficients for the Berry-phase fluctuation mechanism, 
$\alpha_{xy}, \sigma_{xy}  \propto \tau^2 $,
can be also understood
from the following phenomenological argument.
In our mechanism, the Nernst and Hall effects are caused by the asymmetric (or skew) scattering processes of quasiparticles due to chiral superconducting fluctuation.
Contributions from such asymmetric scattering processes to off-diagonal components of transport tensors,
e.g. $\alpha_{xy}$, $\sigma_{xy}$, spin Hall coefficient, and so on,
are proportional to $ \tau^2/\tau_{skew}$, where $\tau$ is the scattering time due to whole scattering processes and $\tau_{skew}$ is that due to asymmetric scattering processes.
This relationship can be derived phenomenologically
by using the Boltzmann equation with the quasiparticle scattering rate that has the asymmetric part \cite{NN_AHE, Sinitsyn}.
In this case, the scattering kernels which cause the skew scattering are not impurities but chiral superconducting fluctuations.
Therefore, $\tau_{skew}$ is independent of the purity of the system,
and then $\alpha_{xy}, \sigma_{xy} \propto \tau^2$, 
in contrast to the usual AHE raised by skew-scattering due to impurities, for which $\tau_{skew} \propto \tau$.

Finally, 
we discuss to what extent our results depend on the form of the 
effective potential, $W({\bm k},\omega_k)$, the spatial dimensionality and pairing symmetry of chiral superconducting states.
We examined that
the $\tau$-dependence of
Eqs. (\ref{final_Nernst_Kubo}) and (\ref{final_Hall}) in clean limit
is not changed by these factors.
However, the magnitude of the transport coefficients depends on the specific form of $W({\bm k},\omega_k)$:
the contributions are decreased as the momentum-dependence of the interaction is stronger.
Also, the most singular parts of $T$-dependence of (\ref{final_Nernst_Kubo}) and (\ref{final_Hall}) are 
not much affected by the specific form of $W({\bm k},\omega_k)$, though the dimensionality may change it.
We present the precise argument in \cite{sup}.

{\it Implications for Experiments} ---
We discuss the implication of our results for experiments.
The Nernst effect is observed by measuring the Nernst coefficient which is the ratio of an induced transverse electric field ($\bm{ E} \parallel \bm{\hat{ y}}$) to product of temperature gradient ($\nabla T  \parallel \bm{ \hat{x}}$) and an applied magnetic field ($\bm{ H}  \parallel \bm{ \hat{z}}$):
$\nu^{NE} = E_y/ (-\nabla_x T ) H = (\alpha_{xy} \sigma_{xx} - \alpha_{xx} \sigma_{xy})/(\sigma^2_{xx}+\sigma^2_{xy})H$.
Usually, the longitudinal conductivity is dominated by contributions from conventional quasiparticles of the Fermi liquid 
rather than that from the superconducting fluctuations, i.e.
$\sigma^{n}_{xx} \gg \sigma^{Fluc}_{xx}$.
Also, for
URu$_2$Si$_2$, as verified experimentally, $\alpha_{xy} / \sigma_{xx} \gg S\tan \Theta_H $, where $S$ is the Seebeck constant and
$\Theta_H$ is the Hall angle \cite{PC_note}. 
Thus, the Nernst coefficient is approximated as $\nu^{NE} \approx \nu^{NE \, n} + \nu^{NE \, Fluc}$,
where $\nu^{NE \, n} $ is the usual Fermi liquid contribution, and $\nu^{NE \, Fluc} = \alpha^{Fluc}_{xy}/ \sigma^n_{xx} H$
with  $\alpha^{Fluc}_{xy}$ the superconducting fluctuation term
(note that what appears in the denominator is not $\sigma^{Fluc}_{xx}$ but $\sigma^{n}_{xx}$).
As mentioned above, $\alpha_{xy}$ due to conventional fluctuation mechanism does not depend on $\tau$, and thus,
$\nu^{NE \, Fluc} \propto \tau ^{-1}$ for non-chiral superconductors, which implies that
 this effect is suppressed for cleaner samples with larger $\tau$ \cite{RullierPRL2006}.
In contrast, the Berry-phase fluctuation mechanism gives $\nu_{BPF}^{NE \, Fluc} \propto \tau ^{1}$ and,
therefore, it is more enhanced for cleaner samples.

Recently, the measurement of the Nernst effect for clean samples of URu$_2$Si$_2$ with different values of  RRRs
was carried out by Kyoto group \cite{PC_note}. 
They found that the Nernst coefficient above $T_c$ is strongly enhanced in cleaner samples.
Therefore, our scenario provides a promising explanation for this anomalous behavior.

{\it Summary} ---
We elucidate the unconventional mechanism of the Nernst and Hall effects raised by Berry-phase fluctuations 
 above $T_c$ in chiral superconductors.
 We propose that our theory can be tested for URu$_2$Si$_2$, which is believed to be a chiral $d+{\rm i}d$ superconductor
 with strong superconducting fluctuations above $T_c$.

We thank T. Yamashita, S. Tonegawa, Y. Tsuruhara, Y. Shimoyama, T. Shibauchi, and, Y. Matsuda, 
for providing experimental data which give us motivation of this study,
 and A. Shitade, K. Shiozaki, T. Nomoto, and R. Ikeda 
 for helpful discussions.
 This work was supported by the Grant-in-Aids for Scientific
Research from MEXT of Japan [Grants No. 23540406, No. 25220711, and No. 25103714 (KAKENHI on Innovative Areas “Topological Quantum Phenomena")]. 
H. S. is supported by a JSPS Fellowship for Young Scientists.

\bibliography{bib_BPF.bib}

\begin{thebibliography}{46}
\bibitem[{com()}]{comment_on_alpha}
\bibinfo{note}{{We used the parameters of the material: $T_c \sim 1.5 {\rm K}$,
  $k_F \sim 0.5 \cdot 10^{10} {\rm m}^{-1}$, $v_F \sim 1 \cdot 10^4 {\rm m/s}$
  \cite{Schlabitz}, and $\Gamma \sim 1.5 {\rm meV}$ \cite{Bourdarot, Wiebe},
  and set the cutoff momentum as $\Lambda = 1/\xi$. We neglected the
  $\tau$-dependence of the transition temperature $T_c$ and coherence length
  $\xi$. These assumptions are justified for $T\tau \gg 1$. In this letter, RRR
  is defined as the ratio of $\sigma_{xx}(T=300{\rm K})$ to
  $\tilde{\sigma}_{xx} (T = 0 {\rm K})$, which is obtain by exptrapolating the
  temperature dependence of the conductivity in the normal-metal region to
  $T=0$ K. It is also assumed that the longitudinal conductivity obeys
  $\sigma_{xx} \propto \tau^{-1} = \tau^{-1}_{imp} + A T^2$, where $\tau_{imp}$
  and $A$ are $T$-independent constant.}}
\end{thebibliography}

\appendix

\renewcommand{\theequation}{S.\arabic{equation}}
\renewcommand{\thefigure}{S.\arabic{figure}}
\renewcommand{\thetable}{S.\arabic{table}}

\renewcommand{\thesection}{}
\renewcommand{\thesubsection}{\Alph{subsection}}
\renewcommand{\thesubsubsection}{ {\rm \thesubsection \arabic{subsubsection}}}

\section*{Supplemental Material}

\subsection{BPS under Magnetic Field}
In this section, we derive the expression of the BPS (3) under a homogeneous magnetic field $(0,0,H)$.

The interaction term of the Hamiltonian (2) for the chirality $C=+1$ channel can be rewritten in real-space representation as:
\begin{eqnarray}
&& H^{int} \nonumber \\
&& = - g
\int d \bm{ r} \left[ V^{+} \left( \frac{-{\rm i} (\partial_{1} - \partial_{2})}{2} , \frac{-{\rm i} (\partial'_{1} - \partial'_{2})}{2} \right)\right. \nonumber \\
&& \qquad \qquad   \left. c^{*}_{\uparrow}(\bm{ r}_1)  c^{*}_{\downarrow}(\bm{ r}_2)   c_{\downarrow}(\bm{ r'}_2)  c_{\uparrow}(\bm{ r'}_1) \right] _{\bm{ r_{1},r_{2},r'_{1},r'_{2}} \to \bm{ r} }
\nonumber \\
&&=  - g
\int d \bm{ r} \left[ \phi \left( \frac{-{\rm i} (\partial_{1} - \partial_{2})}{2} \right) \phi^{\dag} \left( \frac{-{\rm i} (\partial'_{1} - \partial'_{2})}{2} \right) \right. \nonumber \\
&& \qquad \qquad   \left.c^{*}_{\uparrow}(\bm{ r}_1)  c^{*}_{\downarrow}(\bm{ r}_2)  c_{\downarrow}(\bm{ r'}_2)  c_{\uparrow}(\bm{ r'}_1) \right] _{\bm{ r_{1},r_{2},r'_{1},r'_{2}} \to \bm{ r} }. \nonumber \\
\end{eqnarray}
Therefore, the BPS of this channel is given by, 
\begin{eqnarray}
&& \tilde{\Pi}_{C=1}(\bm{ x}, \bm{ y}, \omega_l;H) \nonumber \\
&& =T\sum_{\varepsilon_n}
\left. \left[ \phi \left( \frac{-{\rm i} (\partial_{1} - \partial_{2})}{2} \right) \phi^{\dag} \left( \frac{-{\rm i} (\partial'_{1} - \partial'_{2})}{2} \right) \right. \right. \nonumber \\
&& \left. \left.
\quad \tilde{G}(\bm{ r}'_1, \bm{ r}_1, \varepsilon_{n+l};H)  \tilde{G}(\bm{ r}'_2, \bm{ r}_2, - \varepsilon_n;H) 
\right]
\right|_{ \substack{ \bm{ r}'_1,\bm{ r}'_2 \to \bm{ x}, \\ \bm{ r}_1,\bm{ r}_2 \to \bm{ y}}}. \nonumber \\
 \label{ppc_real}
\end{eqnarray}
In the presence of a uniform magnetic field, 
the one-particle Green function is $\tilde{G}(\bm{ r}', \bm{ r}, \varepsilon_n;H)=e^{-{\rm i}e \Phi(\bm{ r}', \bm{ r})  }G_{core}(\bm{ r}'- \bm{ r}, \varepsilon_n;H)$,
where $\Phi(\bm{ x}, \bm{ y})=   \int_{\bm{x}}^{\bm{y}}  \bm{ A} (\bm{r}) d\bm{ r}$ is an integral of the vector potential 
$\bm{ A} (\bm{r}) $
along a straight line, and
$G_{core}(\bm{ r}'- \bm{ r}, \varepsilon_n;H)$ is the ``core" Green function, which is
translation (then it is a function of $\bm{ r}'- \bm{ r}$), $c$-axis rotation, and gauge invariant \cite{Khodas2003, KarenHall},
and is given as the solution of
\begin{eqnarray}
&&\left[ {\rm i} \varepsilon_n -\frac{1}{2m}(-{\rm i} {\bm \nabla_\rho} + \frac{e}{2} {\bm \rho } \times {\bm H} )^2 + \mu - \hat{\Sigma} \right] G_{core}(\bm{ \rho}, \varepsilon_n;H) \nonumber \\
&&= \delta({\bm \rho}),
\end{eqnarray}
where ${\bm \rho}= {\bm r}' - {\bm r}$, and $\hat{\Sigma}$ is the self-energy that also has the same symmetries as mentioned above.

To proceed further, we take the Landau gauge, $\bm{ A}(\bm{ r})=(0,xH,0)$,
although the final results (\ref{SupBPSundermag}), (\ref{coreBPS}), and (\ref{coreBPSz}) are correct for any gauge choice as shown in the last part of this section.
For this gauge, $\Phi(\bm{ r}', \bm{ r}) =  \frac{eH}{2}(x'+x)(y-y')$.
Now, we introduce the pairing symmetry function of $p_{z}$-wave superconductors, $\phi^{z} ({\bm k}) = \sqrt{3} k_{z}/k_{F}$, and
let $\theta= -{\rm i}e( \Phi(\bm{ r}'_1, \bm{ r}_1) + \Phi(\bm{ r}'_2, \bm{ r}_2))$, $\phi_{12} = \phi \left( \frac{-{\rm i} (\partial_{1} - \partial_{2})}{2} \right)$, 
and $\phi^z_{12} = \phi^z \left( \frac{-{\rm i} (\partial_{1} - \partial_{2})}{2} \right)$.
The commutatiors of $\phi$ and $\theta$ are given by
$[\phi_{12}, \theta]= \frac{1}{4} \sqrt{\frac{5}{2}}\frac{{\rm i}eH}{k_F} (-x_1-x_1' +x_2 +x_2' + {\rm i}y_1 -{\rm i}y_1' -{\rm i} y_2' +{\rm i} y_2' ) \phi^z_{12} $,
$[\phi^{\dag}_{1'2'}, \theta]= \frac{1}{4} \sqrt{\frac{5}{2}}\frac{{\rm i}eH}{k_F} (-x_1-x_1' +x_2 +x_2' + {\rm i}y_1 -{\rm i}y_1' -{\rm i} y_2' +{\rm i} y_2' ) \phi^z_{1'2'} $, and
$[\phi_{12}, [\phi^{\dag}_{1'2'}  \theta]]= -\frac{5eH}{4k_F^2}  \phi^z_{12} \phi^{z}_{1'2'} $.
Then,  $[\phi_{12}, \theta] \to 0$ and $[\phi^{\dag}_{1'2'}, \theta] \to 0$ as $\bm{ r}'_1,\bm{ r}'_2 \to \bm{ x}$ and $\bm{ r}_1,\bm{ r}_2 \to \bm{ y}$.
By using these commutation relations and the formulae $e^{B}Ae^{-B}=A+[B,A]+[B,[B,A]]+ \cdots $ and $[PQ, R]=P[Q,R]+[P,R]Q$, we obtain
\begin{eqnarray}
&& \tilde{\Pi}_{C=1}(\bm{ x}, \bm{ y}, \omega_q;H) \nonumber \\
&&=e^{-2{\rm i}e \Phi(\bm{ x}, \bm{ y}) }  \nonumber \\
&&\left[
T\sum_{\varepsilon_n}
{\scriptstyle \phi_{12} \phi^{\dag}_{1'2'} 
G_{core}(\bm{ r}'_1- \bm{ r}_1, \varepsilon_{n+q};H)  G_{core}(\bm{ r}'_2- \bm{ r}_2, - \varepsilon_n;H)} \right. \nonumber \\
&& -\frac{5eH}{4k_F^2} T\sum_{\varepsilon_n} {\scriptstyle  \phi^z_{12} \phi^{z}_{1'2'} G_{core}(\bm{ r}'_1- \bm{ r}_1, \varepsilon_{n+q};H)   }\nonumber \\
&&\qquad \qquad  \qquad {\scriptstyle \left. G_{core}(\bm{ r}'_2- \bm{ r}_2, - \varepsilon_n;H) \right]_{ \substack{  \bm{ r}'_1,\bm{ r}'_2 \to \bm{ x} ,\\ \bm{ r}_1,\bm{ r}_2 \to \bm{ y}} }} \nonumber \\
&&= e^{-{\rm i}2 e \Phi(\bm{ x}, \bm{ y})  } \nonumber \\
&& \left[ \Pi (\bm{ x-y}, \omega_{q} ;H) -   \frac{5eH}{4k^2_{F}} \Pi' (\bm{ x-y}, \omega_{q} ;H) \right] , \nonumber \\
\label{SupBPSundermag}
\end{eqnarray}
where 
\begin{eqnarray}
&&\Pi (\bm{ \rho}, \omega_{q} ;H)  \nonumber \\
&&= \left[ T\sum_{\varepsilon_n} |\phi|^2 \left( \frac{-{\rm i} (\partial_{{\bm \rho_1}}  -\partial_{{\bm \rho_2}} ) }{2} \right)  G_{core}({\bm \rho}_1, \varepsilon_{n+q};H) \right. \nonumber \\
&&\qquad \qquad   \left. G_{core}({\bm \rho}_2, - \varepsilon_n; H) \right]_{ {\bm \rho}_1, {\bm \rho}_2 \to {\bm \rho}}, \label{coreBPS} \\
&& \Pi' (\bm{ \rho}, \omega_{q} ;H)  \nonumber \\
&&= \left[  T\sum_{\varepsilon_n} |\phi^z|^2 \left( \frac{-{\rm i} (\partial_{{\bm \rho_1}}  -\partial_{{\bm \rho_2}} ) }{2} \right)  G_{core}({\bm \rho}_1, \varepsilon_{n+q};H) \right. \nonumber \\
&&\qquad \qquad \left.  G_{core}({\bm \rho}_2, - \varepsilon_n; H) \right]_{ {\bm \rho}_1, {\bm \rho}_2 \to {\bm \rho}} \label{coreBPSz}
\end{eqnarray}
are ``core" BPSs, which preserve spatial translation, $c$-axis rotation, and gauge invariances.
Here $|\phi({\bm k})|^2 = 15k_z^2(k_x^2+k_y^2)/2k_F^4  $ and $|\phi^z({\bm k})|^2=3k_z^2/k_F^2$.
Then, we arrive at Eq.(3) for $C=+1$.
Carrying out similar calculations with the effective interaction term $V^{-}$, we can obtain Eq.(3) for $C=-1$.

So far we have used the Landau gauge.
However,  by using the formulae, $\left. [\phi_{12},e^{-{\rm i}(\chi({\bm r_1})+\chi({\bm r_2))}}] \right|_{{\bm r}_1,{\bm r}_2 \to {\bm y}}=0$,
and so on, with $\chi$ an arbitrary function,
we can prove that
the final results, (\ref{SupBPSundermag}), (\ref{coreBPS}), and (\ref{coreBPSz}), are gauge invariant.

\subsection{Nernst and Hall Conductivities}
\label{Derivations_of_Nernst_and_Hall}

In this section, we present the derivation of the Nernst and Hall conductivities 
for the ANE and AHE caused by the Berry-phase fluctuation mechanism,
(4) and (5).

Formally, to obtain the whole contributions from the superconducting fluctuations to the Kubo term of the Peltier coefficient and the Hall coefficient,
one has to evaluates all possible Feynman diagrams
that consist of two current vertices, superconducting fluctuation propagators, $\tilde{L}_C$, and
Green functions, $\tilde{G}$, in a magnetic field.
We focus on the case with a weak magnetic field, where the Nernst and Hall conductivities are linear in $H$.
Up to the linear order of $H$, the fluctuation propagator (3) is divided into two term:
\begin{eqnarray}
\tilde{L}_C=\tilde{L}_0 + \tilde{L}' _C+ \mathcal{O}(H^2) ,
\label{tildLC}
\end{eqnarray}
where 
\begin{eqnarray}
&&\tilde{L}^{-1}_0({\bm x},{\bm y},\omega_q;H) \nonumber \\
&&= - \frac{\delta({\bm x}-{\bm y})}{g} + e^{-{\rm i}2 e \Phi(\bm{ x}, \bm{ y})  } \Pi({\bm x}-{\bm y},\omega_q;H),
\end{eqnarray}
is the conventional part of the fluctuation propagator,
and
\begin{eqnarray}
\tilde{L}' _C({\bm x}-{\bm y},\omega_q;H) = C\cdot \left( \frac{ 5eH}{4k_F^2g}\right)  \left[ L({\bm x}-{\bm y},\omega_q)\right]^2,
\label{chiraliydepL}
\end{eqnarray}
is the chirality-dependent one which is characteristic of chiral superconductors.
Here $L({\bm x}-{\bm y},\omega_q)$ is the fluctuation propagator of $d_{zx} \pm {\rm i} d_{zy}$-wave channel in zero magnetic field.
The concrete expression of $L({\bm x}-{\bm y},\omega_q)$ and  its derivation are given in
Sec.\ref{FPunderzeromag}. From Eq.(\ref{tildLC}), we see
that in the linear order of $H$, the whole contributions to the Nernst and Hall conductivities
are separated into two parts:
[A] contribution from diagrams which 
do not include the chirality-dependent fluctuation propagator, $\tilde{L}'_C$,
[B] contribution from diagrams which include one chirality-dependent fluctuation propagator $\tilde{L}'_C$.

The contribution [A] also appears for the case of non-chiral superconductors, such as $s$- and $d_{x^2-y^2}$-wave pairings,
which can be described by conventional theories \cite{Larkin,Ussishkin2003, Serbyn2009},
and, therefore, physically, this part corresponds to the contributions due to Lorentz force on quasiparticles and fluctuating Cooper pairs.
On the other hand, [B] is unique to chiral superconductors, raised by chirality-polarization, and
as shown below,
associated with the ANE and AHE caused by the Berry-phase fluctuation mechanism without Lorentz force.

From now on, we concentrate on the latter contribution, and 
write down the correlation functions for the Nernst and Hall conductivities in the form,  
\begin{eqnarray}
A_{xy}(\omega_l)=-2eT \sum_{{\bm q}, \omega_q, C=\pm 1} \tilde{L}'_C({\bm q}, \omega_q;H) \bar{A}_C ({\bm q}, \omega_q; \omega_l) , \nonumber \\
\end{eqnarray}
where only the odd part of $\bar{A}_C$ with respect to time reversal: $C \to -C$, gives nonzero contributions,
since $\tilde{L}'_C$ is odd.

\begin{figure}[htbp]
 \includegraphics[width=80mm]{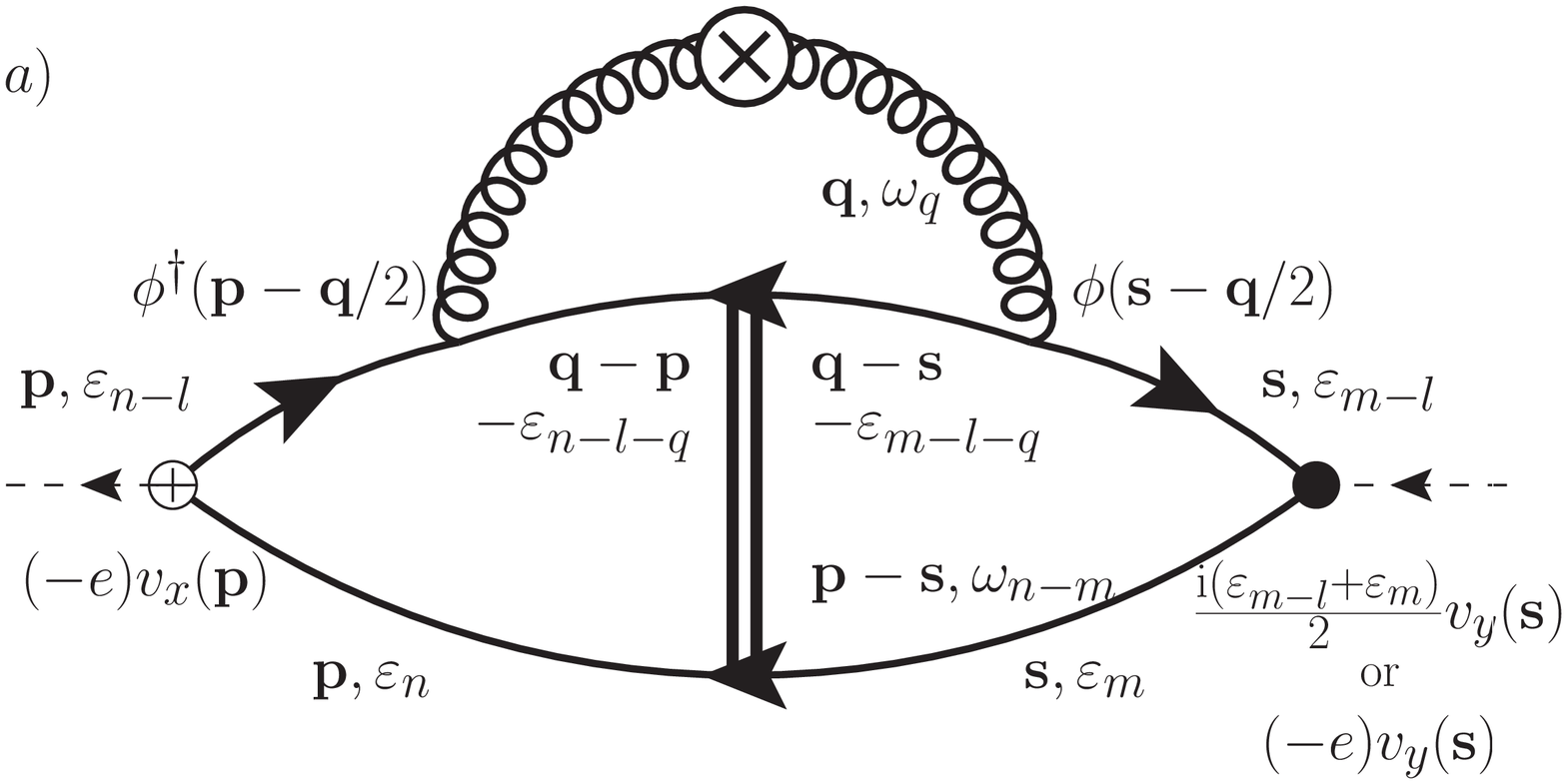}\\ 
 \includegraphics[width=80mm]{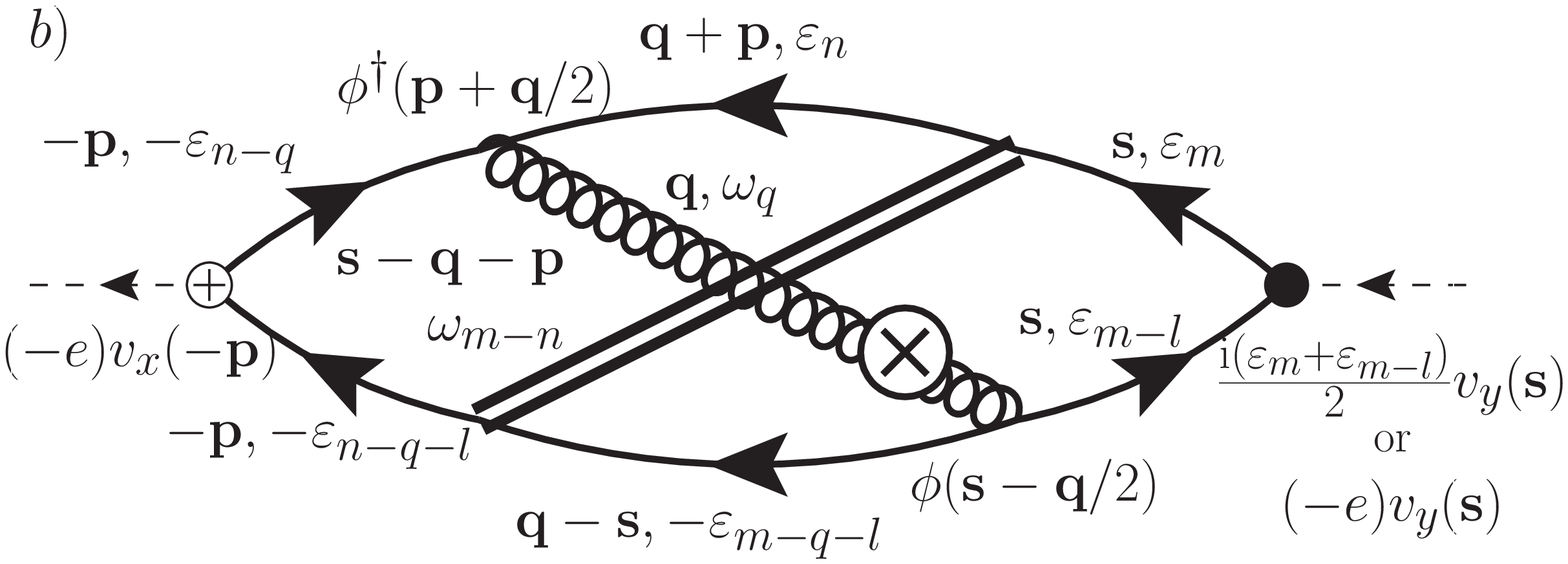}\\ 
 \includegraphics[width=80mm]{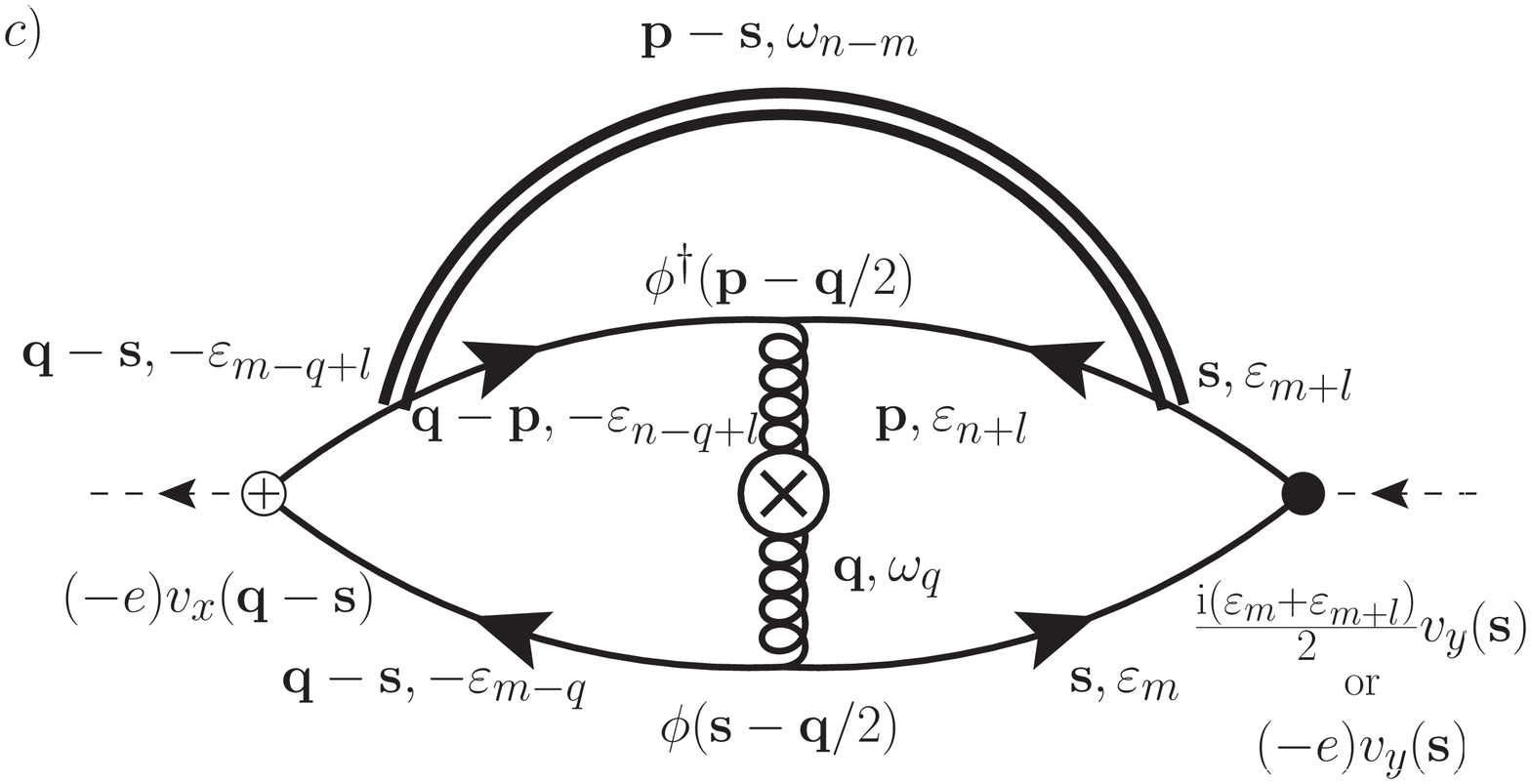}
 \centering
 \caption{Diagrams $a)$, $b)$, and $c)$ of FIG. 2 with wave numbers and frequencies explicitly shown.}          
 \label{new_with_wave_number}
 \end{figure}

We examine the leading-order diagrams belonging to [B], which give the dominant contribution. 
It is found that the three diagrams which give leading-order contributions in conventional theories, i.e.
the Aslamazov-Larkin (AL), Maki-Thompson (MT), and density-of-states (DOS) diagrams (upper panel in FIG. 1)
do not contribute in this case,
and generally, all contributions from diagrams belonging to the classes of the lower panel in FIG. 1 are zero.
The reason is that in these diagrams the paring function appear as $|\phi |^2$,
and, therefore, their contributions to $\bar{A}_C$ have only even part with respect to time reversal: 
\begin{eqnarray}
\bar{A}^{AL,MT,DOS}_{C}  = \bar{A}^{AL,MT,DOS}_{-C},
\end{eqnarray}
which do not contribute to the correlation function as mentioned above.
The lowest order diagrams which do not belong to these classes and give nonzero contributions are depicted in FIG. 2 in the main text.
For clarity, we depicted them more explicitly in FIGs. \ref{new_with_wave_number}$a$-$c)$).
In these diagrams, a renormalized four-point vertex, $W({\bm k}, \omega_k)$, (double line) raised by electron-electron interaction is inserted.
To proceed further, we introduce a simple model: $W({\bm k}, \omega_k)=W_0/(1+|\omega_k|/\Gamma)$,
the four-point vertex mediated via the short-range spin fluctuation,
where we discuss about the assumption in Sec. \ref{From_of_W}.

Now, we calculate contributions from diagrams $a)$, $b)$, $c)$, and their mirror images in FIG. 2 
to the Kubo terms of the Nernst conductivity, $\alpha_{xy}^{i) \, Kubo}$, ($i=a$, $b$, and $c$), and the Hall conductivity, $\sigma^{i)}_{xy}$, ($i=a$, $b$, and $c$).
They are, respectively, given by
\begin{eqnarray}
\alpha_{xy}^{i) \, Kubo}=\left.\frac{1}{T}\frac{A^{i) \, R}_{\alpha\beta}(\omega)}{(-i\omega)}\right|_{\omega\rightarrow 0},
\end{eqnarray}
\begin{eqnarray}
\sigma^{i)}_{xy}=\left.\frac{S^{i) \, R}_{\alpha\beta}(\omega)}{(-i\omega)}\right|_{\omega\rightarrow 0}, 
\end{eqnarray}  
where $A^{i) \, R}_{\alpha\beta}(\omega)$ is the retarded correlation function of heat currents, and
$S^{i) \, R}_{\alpha\beta}(\omega)$ is that of charge currents.
Their corresponding correlation functions with the Matsubara frequencies are, respectively, 
\begin{eqnarray}
A^{i)}_{xy} (\omega_l) &=& -2eT \sum_{\omega_q,\bm{ q},C= \pm 1} \tilde{L}'_C(\bm{ q},\omega_{q}; H) \bar{A}^{i)}_C(\bm{ q},\omega_{q};\omega_{l}), \nonumber \\
&&\label{corr_A} \\
S^{i)}_{xy} (\omega_l) &=& 2e^2T \sum_{\omega_q,\bm{ q},C= \pm 1} \tilde{L}'_C(\bm{ q},\omega_{q}; H) \bar{S}^{i)}_C(\bm{ q},\omega_{q};\omega_{l}), \nonumber \\
&& \label{corr_S} 
\end{eqnarray}
where $\omega_q$ and $\omega_l$ are the Matsubara frequencies, and
$\bm{q}$ is a wave number. Here,
\begin{eqnarray}
&&\bar{A}^{i)}_{C=1}(\bm{ q}, \omega_{q}; \omega_l)  \nonumber \\
&&= -2T^2 {\rm Re} \left[ \sum_{n,m} X^{i)}(\bm{ q}, n,\omega_{q}; \omega_l) Y^{i)}(\bm{ q}, m,\omega_{q}; \omega_l) \right. \nonumber \\
&&\qquad \qquad \quad \left. \frac{W_0}{1+|\omega_{n-m}|/\Gamma} \right] , \label{barAdef} \\ 
&&\bar{S}^{i)}_{C=1}(\bm{ q}, \omega_{q}; \omega_l)  \nonumber \\
&&=-2T^2 {\rm Re} \left[ \sum_{n,m} X^{i)}(\bm{ q}, n,\omega_{q}; \omega_l) Z^{i)}(\bm{ q}, m,\omega_{q}; \omega_l)  \right. \nonumber \\
&& \qquad \qquad \quad \left. \frac{W_0}{1+|\omega_{n-m}|/\Gamma} \right],
\end{eqnarray}
and
\begin{eqnarray}
&&X^{a)}(\bm{ q}, n,\omega_q; \omega_l) \nonumber \\
&&=\sum_{\bm{ p}}  \phi^{\dag} (\bm{ p}-\frac{\bm{ q}}{2}) G(\bm{ q-p}, - \varepsilon_{n-l-q}) G(\bm{ p}, \varepsilon_{n-l}) \nonumber \\
&& \quad G(\bm{ p}, \varepsilon_{n}) v_x (\bm{ p}) , \label{Xadef}\\
&&Y^{a)}(\bm{ q}, m,\omega_q; \omega_l)\nonumber \\
&&=\sum_{\bm{ s}} \phi (\bm{ s}-\frac{\bm{ q}}{2}) G(\bm{ q-s}, - \varepsilon_{m-l-q}) G(\bm{ s},  \varepsilon_{m-l})\nonumber\\
&& \quad  G(\bm{ s}, \varepsilon_{m}) \frac{{\rm i}(\varepsilon_{m-l} + \varepsilon_{m})}{2} v_y (\bm{ s}) , \\
&&Z^{a)}(\bm{ q}, m,\omega_q; \omega_l)\nonumber \\ 
&&= \sum_{\bm{ s}} \phi (\bm{ s}-\frac{\bm{ q}}{2}) G(\bm{ q-s}, - \varepsilon_{m-l-q}) G(\bm{ s},  \varepsilon_{m-l}) \nonumber \\
&& \quad G(\bm{ s}, \varepsilon_{m})  v_y (\bm{ s}) ,
\end{eqnarray}
\begin{eqnarray}
&&X^{b)}(\bm{ q}, n,\omega_q; \omega_l) \nonumber \\
&& = \sum_{\bm{ p}}   \phi^{\dag} (\bm{ p}+\frac{\bm{ q}}{2}) G(\bm{ q+p},  \varepsilon_{n}) G(-\bm{ p},-\varepsilon_{n-q}) \nonumber \\
&& \quad  G(-\bm{ p},- \varepsilon_{n-q-l})  v_x (-\bm{ p}),  \\
&&Y^{b)}(\bm{ q}, m,\omega_q; \omega_l)\nonumber \\
&& =\sum_{\bm{ s}} \phi (\bm{ s}-\frac{\bm{ q}}{2}) G(\bm{ q-s}, - \varepsilon_{m-q-l}) G(\bm{ s}, \varepsilon_{m-l}) \nonumber \\
&& \quad  G(\bm{ s}, \varepsilon_{m}) \frac{{\rm i}(\varepsilon_m + \varepsilon_{m-l})}{2} v_y (\bm{ s}), \\
&&Z^{b)}(\bm{ q}, m,\omega_q; \omega_l) \nonumber \\
&&=\sum_{\bm{ s}} \phi (\bm{ s}-\frac{\bm{ q}}{2}) G(\bm{ q-s}, - \varepsilon_{m-q-l}) G(\bm{ s}, \varepsilon_{m-l}) \nonumber \\
&& \quad G(\bm{ s}, \varepsilon_{m}) v_y (\bm{ s}),
\end{eqnarray}
\begin{eqnarray}
&&X^{c)}(\bm{ q}, n,\omega_q; \omega_l) \nonumber \\
&& =\sum_{\bm{ p}} \phi^{\dag} (\bm{ p}-\frac{\bm{ q}}{2}) G(\bm{ q-p}, - \varepsilon_{n+l-q}) G(\bm{ p},\varepsilon_{n+l}), \\
&&Y^{c)}(\bm{ q}, m,\omega_q; \omega_l) \nonumber \\
&& =\sum_{\bm{ s}}  \phi (\bm{ s}-\frac{\bm{ q}}{2})  G(\bm{ q}-\bm{ s}, -\varepsilon_{m-q})  G(\bm{ s},  \varepsilon_{m}) G(\bm{ s}, \varepsilon_{m+l})  \nonumber\\
&&\quad G(\bm{ q}- \bm{ s}, - \varepsilon_{m+l-q})   v_x (\bm{ q}-\bm{ s})  \frac{{\rm i}(\varepsilon_{m} + \varepsilon_{m+l})}{2} v_y (\bm{ s}),  \nonumber \\
&& \\
&&Z^{c)}(\bm{ q}, m,\omega_q; \omega_l) \nonumber \\
&& = \sum_{\bm{ s}}  \phi (\bm{ s}-\frac{\bm{ q}}{2})  G(\bm{ q}-\bm{ s}, -\varepsilon_{m-q})  G(\bm{ s},  \varepsilon_{m}) G(\bm{ s}, \varepsilon_{m+l}) \nonumber \\
&& \quad G(\bm{ q}- \bm{ s}, - \varepsilon_{m+l-q})   v_x (\bm{ q}-\bm{ s})   v_y (\bm{ s}).  \label{Zdef}
\end{eqnarray} 
We can also obtain the expressions for
$\bar{A}^{i)}_{C=-1}$ and $\bar{S}^{i)}_{C=-1}$ 
by using chirality-inversion (time-reversal) transformation, $\phi \to\phi^{\dag}$ and $\phi^{\dag} \to\phi$ in $X^{i)}$, $Y^{i)}$, and $Z^{i)}$, 
of Eqs. (\ref{barAdef}-\ref{Zdef}).
In the above equations, the one-particle Green function, chiral $d_{zx} + {\rm i} d_{zy}$-wave pairing symmetry function, and velocity of quasiparticles are defined as
$G({\bm k}, \varepsilon_{k})^{-1}={\rm i}\tilde{\varepsilon}_k -\xi_{{\bm k}}={\rm i} ( \varepsilon_{k} + {\rm sgn  (\varepsilon_{k} )/2\tau}) -\xi_{{\bm k}} $,
$\phi(\bm{ k})= \sqrt{15/2} k_z (k_x + {\rm i} k_y)/k^2_{F}$, and 
${\bm v}({\bm k}) = \partial  \xi_{{\bm k}} / \partial {\bm k}$, respectively.
Here $\tau$ is the relaxation time of quasiparticles.
For simplicity, we take the spherical Fermi surface: $ \xi_{{\bm k}} = k^2 /2m - \mu$,
where $m$ and $\mu$ are the mass and chemical potential of the quasiparticles, respectively.

We, henceforth, neglect quantum superconducting fluctuations keeping only terms with $\omega_q =0$.
Since singular contributions at $T_c$ come from long-wave length regions where the center-of-mass momentum of fluctuating Cooper pairs $q$ is small,
we concentrate on the analysis of $\tilde{A}^{i)}$ and $\tilde{S}^{i)}$ for small $q$.
Then, expanding $X^{i)}$, $Y^{i)}$, and $Z^{i)}$ as power series of $q/k_F$, we obtain 
 
\begin{eqnarray}
&&X^{a)}(\bm{q}, n, \omega_q=0; \omega_l) \nonumber \\
&& = N(0) q_{z} \left[
{\rm i}\pi \sqrt{ \frac{5}{24}} \frac{v_{F}}{ k_{F}} \frac{{\rm sgn}(n + 1/2)}{(|\tilde{\varepsilon}_{n}|+|\tilde{\varepsilon}_{n-l}|) |\tilde{\varepsilon}_{n-l}|} \right. \nonumber \\
&&\quad \left. - \frac{\pi}{2} \sqrt{\frac{1}{30}} v^{2}_{F} 
\left\{ 
 \begin{array}{l}
  {\displaystyle \frac{|\tilde{\varepsilon}_{n}|+3|\tilde{\varepsilon}_{n-l}|}{|\tilde{\varepsilon}_{n-l}|^2(|\tilde{\varepsilon}_{n}|+|\tilde{\varepsilon}_{n-l}|)^2} ,} \\ 
  \quad  { \scriptstyle {\rm for} \, (n-l+1/2)(n+1/2)>0}\\
  {\displaystyle \frac{-1}{|\tilde{\varepsilon}_{n-l}|^2(|\tilde{\varepsilon}_{n}|+|\tilde{\varepsilon}_{n-l}|)} , }\\
   \quad { \scriptstyle  {\rm for} \, (n-l+1/2)(n+1/2)<0}\\
    \end{array}
\right\}
 \right] \nonumber \\
 && \quad + \mathcal{O} \left( \left( \frac{q}{k_{F}} \right)^2 \right), \\
&&Y^{a)}(\bm{q}, m, \omega_q=0; \omega_l) \nonumber \\
&& =- \frac{(\varepsilon_{m-l}+\varepsilon_{m})}{2} X^{a)}(q, m, \omega_q=0; \omega_l),  \\ 
&&Z^{a)}(\bm{q}, m, \omega_q=0; \omega_l) \nonumber \\
&& = {\rm i} X^{a)}(q, m, \omega_q=0; \omega_l), 
\end{eqnarray}
\begin{eqnarray}
&&X^{b)}(\bm{q}, n, \omega_q =0; \omega_l) \nonumber \\
&&= N(0) q_{z} \left[
- {\rm i}\pi \sqrt{ \frac{5}{24}} \frac{v_{F}}{ k_{F}} \frac{{\rm sgn}(n -l + 1/2)}{(|\tilde{\varepsilon}_{n}|+|\tilde{\varepsilon}_{n-l}|) |\tilde{\varepsilon}_{n}|} \right. \nonumber \\
&& \quad \left. - \frac{\pi}{2} \sqrt{\frac{1}{30}} v^{2}_{F} 
\left\{ 
 \begin{array}{l}
  {\displaystyle \frac{|\tilde{\varepsilon}_{n-l}|+3|\tilde{\varepsilon}_{n}|}{|\tilde{\varepsilon}_{n}|^2(|\tilde{\varepsilon}_{n-l}|+|\tilde{\varepsilon}_{n}|)^2} , }\\
   \quad { \scriptstyle {\rm for} \, (n-l+1/2)(n+1/2)>0}\\
   {\displaystyle \frac{-1}{|\tilde{\varepsilon}_{n}|^2(|\tilde{\varepsilon}_{n-l}|+|\tilde{\varepsilon}_{n}|)}, }\\ 
   \quad { \scriptstyle {\rm for} \, (n-l+1/2)(n+1/2)<0}\\
    \end{array}
\right\}  \right] \nonumber \\
&& \quad + \mathcal{O} \left( \left( \frac{q}{k_{F}} \right)^2 \right),
\end{eqnarray}
\begin{eqnarray}
&&Y^{b)}(\bm{q}, m, \omega_q=0; \omega_l) \nonumber \\
&& = N(0) q_{z} \frac{\varepsilon_m + \varepsilon_{m-l}}{2} \left[
- {\rm i}\pi \sqrt{ \frac{5}{24}} \frac{v_{F}}{ k_{F}} \frac{{\rm sgn}(m+ 1/2)}{(|\tilde{\varepsilon}_{m}|+|\tilde{\varepsilon}_{m-l}|) |\tilde{\varepsilon}_{m-l}|} \right. \nonumber \\
&& \quad \left. + \frac{\pi}{2} \sqrt{\frac{1}{30}} v^{2}_{F} 
\left\{ 
 \begin{array}{l}
  {\displaystyle \frac{|\tilde{\varepsilon}_{m}|+3|\tilde{\varepsilon}_{m-l}|}{|\tilde{\varepsilon}_{m-l}|^2(|\tilde{\varepsilon}_{m}|+|\tilde{\varepsilon}_{m-l}|)^2} ,}\\
   \quad { \scriptstyle {\rm for} \, (m-l+1/2)(m+1/2)>0} \\
 {\displaystyle  \frac{-1}{|\tilde{\varepsilon}_{m-l}|^2(|\tilde{\varepsilon}_{m}|+|\tilde{\varepsilon}_{m-l}|)} ,} \\
  \quad { \scriptstyle {\rm for} \, (m-l+1/2)(m+1/2)<0} \\
    \end{array}
\right\}  \right] \nonumber \\
&& \quad + \mathcal{O} \left( \left( \frac{q}{k_{F}} \right)^2 \right), 
\end{eqnarray}
\begin{eqnarray}
&&Z^{b)}(\bm{q}, m, \omega_q=0; \omega_l) \nonumber \\
&& = N(0) q_{z}  \left[
- \pi \sqrt{ \frac{5}{24}} \frac{v_{F}}{ k_{F}} \frac{{\rm sgn}(m+ 1/2)}{(|\tilde{\varepsilon}_{m}|+|\tilde{\varepsilon}_{m-l}|) |\tilde{\varepsilon}_{m-l}|} \right. \nonumber \\
&& \quad \left. - {\rm i} \frac{\pi}{2} \sqrt{\frac{1}{30}} v^{2}_{F} 
\left\{ 
 \begin{array}{l}
  {\displaystyle \frac{|\tilde{\varepsilon}_{m}|+3|\tilde{\varepsilon}_{m-l}|}{|\tilde{\varepsilon}_{m-l}|^2(|\tilde{\varepsilon}_{m}|+|\tilde{\varepsilon}_{m-l}|)^2 ,}} \\
  { \scriptstyle \quad {\rm for} \, (m-l+1/2)(m+1/2)>0} \\
  {\displaystyle \frac{-1}{|\tilde{\varepsilon}_{m-l}|^2(|\tilde{\varepsilon}_{m}|+|\tilde{\varepsilon}_{m-l}|)} ,} \\
  { \scriptstyle \quad {\rm for} \, (m-l+1/2)(m+1/2)<0}\\
    \end{array}
\right\}  \right] \nonumber \\
&& \quad + \mathcal{O} \left( \left( \frac{q}{k_{F}} \right)^2 \right),  
\end{eqnarray}
\begin{eqnarray}
&&X^{c)}(\bm{q}, n, \omega_q=0; \omega_l) =  \mathcal{O} \left( \left( \frac{q}{k_{F}} \right)^2 \right), \\
&&Y^{c)}(\bm{q}, m, \omega_q=0; \omega_l) =  \mathcal{O} \left( \frac{q}{k_{F}}  \right), \\
&&Z^{c)}(\bm{q}, m, \omega_q=0; \omega_l) =  \mathcal{O} \left( \frac{q}{k_{F}}  \right),
\end{eqnarray}
for $\omega_l>0$, where $N(0)$ is the density-of-states at the Fermi surface.
Since $\bar{A}^{c)}_{C=1}(\bm{ q}, \omega_q=0; \omega_l)= \mathcal{O} \left( \left( \frac{q}{k_{F}} \right)^3 \right)$, and $ \bar{S}^{c)}_{C=1}(\bm{ q}, \omega_q=0; \omega_l) = \mathcal{O} \left( \left( \frac{q}{k_{F}} \right)^3 \right)$ , the contribution from the $c)$ diagram is less singular near $T_c$ than that from $a)$ or $b)$. Therefore, we neglect the contributions from this diagram.

The most singular part in DC limit for a clean system with large $\tau$ arises from the summation over $n$ in the region $n,m=0,1,...,l-1$, where $|\tilde{\varepsilon}_{n}|+|\tilde{\varepsilon}_{n-l}|=\omega_l + 1/\tau$.
We take only such terms and obtain
\begin{eqnarray}
&&\bar{A}^{a)}_{C=1}(\bm{ q}, \omega_q=0; \omega_l) \nonumber \\
&& =    \frac{N(0)^2  W_0 q^2_z}{( \omega_l +1/\tau)^2} \sum_{0 \leq n,m \leq l-1} \frac{1}{1+(2\pi T/\Gamma)|n-m|}  \nonumber \\
&&\quad \times \left\{ \mathsmaller{ - \frac{5 \pi  }{48} \frac{T v^2_F}{k^2_F}  \frac{2m-l+1}{(n-l+1/2-1/4\pi \tau T)(m-l+1/2-1/4\pi \tau T)} }\right. \nonumber \\
&&\quad \mathsmaller{ \left. + \frac{1}{960 \pi}  \frac{v^4_F}{T} \frac{2m-l+1}{(n-l+1/2-1/4\pi \tau T)^2(m-l+1/2-1/4\pi \tau T)^2} \right\} }  
,    \nonumber \\
&&  \label{Aa_til} 
\end{eqnarray}
\begin{eqnarray}
&&\bar{A}^{b)}_{C=1}(\bm{ q}, \omega_q= 0; \omega_l)  \nonumber \\
&& =   \frac{N(0)^2  W_0  q^2_z}{( \omega_l +1/\tau)^2} \sum_{0 \leq n,m \leq l-1} \frac{1}{1+(2\pi T/\Gamma)|n-m|} \nonumber \\
&& \quad \times \mathsmaller{\left\{ - \frac{5 \pi  }{48} \frac{T v^2_F}{k^2_F}  \frac{2m-l+1}{(n+1/2+1/4\pi \tau T)(m-l+1/2-1/4\pi \tau T)} \right. } \nonumber \\
&&\quad \mathsmaller{ \left. + \frac{1}{960 \pi }  \frac{v^4_F}{T}  \frac{2m-l+1}{(n+1/2+1/4\pi \tau T)^2(m-l+1/2-1/4\pi \tau T)^2} \right\} } 
 ,     \nonumber \\
&& \label{Ab_til}  \\
\, \nonumber 
\end{eqnarray}
\begin{eqnarray}
&& \bar{S}^{a)}_{C=1}(\bm{ q}, \omega_q=0; \omega_l) \nonumber \\
&& =    \frac{N(0)^2  W_0 q^2_z}{( \omega_l +1/\tau)^2}  \sum_{0 \leq n,m \leq l-1} \frac{1}{1+(2\pi T/\Gamma)|n-m|}\nonumber \\
&& \quad \times \mathsmaller{  \frac{-1}{48 \pi } \frac{v^3_F}{k_F T}  \frac{1}{(n-l+1/2-1/4\pi \tau T)(m-l+1/2-1/4\pi \tau T)^2}},  \nonumber \\
&& \quad   \label{Sa_til} 
\end{eqnarray}
\begin{eqnarray}
&& \bar{S}^{b)}_{C=1}(\bm{ q}, \omega_q=0; \omega_l) \nonumber \\
&&  =  \frac{ N(0)^2  W_0 q^2_z}{( \omega_l +1/\tau)^2}  \sum_{0 \leq n,m \leq l-1} \frac{1}{1+(2\pi T/\Gamma)|n-m|}  \nonumber \\
&&\quad \mathsmaller{ \times \frac{1}{48 \pi }  \frac{v^3_F}{k_F T} \frac{1}{(n+1/2+1/4\pi \tau T)(m-l+1/2-1/4\pi \tau T)^2} } . \nonumber \\
&&  \label{Sb_til}
\end{eqnarray}
Up to now, we have calculated only the $C=1$ terms.
However, by carrying out calculations similar to (\ref{Xadef}-\ref{Sb_til}), we immediately find,
\begin{eqnarray}
\bar{A}^{i)}_{C=-1}(\bm{ q}, \omega_q= 0; \omega_l) = - \bar{A}^{i)}_{C=1}(\bm{ q}, \omega_q= 0; \omega_l), \\
\bar{S}^{i)}_{C=-1}(\bm{ q}, \omega_q= 0; \omega_l) = - \bar{S}^{i)}_{C=1}(\bm{ q}, \omega_q= 0; \omega_l).
\end{eqnarray}

Now, we introduce $\bar{\alpha}^{i)}_C(\bm{ q}),\bar{\sigma}^{i)}_C(\bm{ q})$ ($i=a$, $b$, and $c$) defined as, 
\begin{eqnarray}
&& \bar{\alpha}^{i)}_{C}(\bm{ q}) = \frac{1}{T} \left. \frac{\bar{A}^{i)}_C(\bm{ q}, \omega_q=0; \omega_l)}{\omega_l} \right|_{{\rm i} \omega_l \to \omega + {\rm i} 0}, \nonumber \\
&& \bar{\sigma}^{i)}_{C}(\bm{ q}) = \left. \frac{\bar{S}^{i)}_C(\bm{ q}, \omega_q=0; \omega_l)}{\omega_l} \right|_{{\rm i} \omega_l \to \omega + {\rm i} 0}.
\end{eqnarray}
Then, we obtain
\begin{eqnarray}
&& \bar{\alpha}^{a)}_C(\bm{ q})  \nonumber \\
&& = C \cdot N(0)^2  W_0 \tau^2 q^2_z \times  \nonumber \\
&&\quad \left[  \mathsmaller{  \frac{5\{ \pi^2 u_{(1,0)}(t,\gamma) -14 \zeta(3) \gamma u_{(1,1)} (t,\gamma) \}}{96} \frac{v_F^2}{k_F^2 T } } \right. \nonumber  \\
&& \quad \left.  \mathsmaller{ + \frac{\{- \pi^4 u_{(2,1)}(t,\gamma)  + (- \psi^{(4)}(1/2)/6) \gamma u_{(2,2)} (t,\gamma) \} }{1920 \pi^2} \frac{v_F^4}{T^3} }\right] , \nonumber \\
&& \label{eq:al1} \\
&& \bar{\alpha}^{b)}_C(\bm{ q}) \nonumber \\
&& = C \cdot N(0)^2  W_0 \tau^2 q^2_z  \nonumber \\
&& \quad \left[ \mathsmaller{ \frac{5\{ - \pi^2 u_{(1,0)}(t,\gamma) + \pi^2 \gamma w_{(1,1)} (t,\gamma) \}}{96} \frac{v_F^2}{k_F^2 T} } \right. \nonumber \\
&& \quad \left. + \mathsmaller{  \frac{\{- 2\pi^2 w_{(2,1)}(t,\gamma) +2 \pi^2 \gamma w_{(2,2)} (t,\gamma) \} }{1920 \pi^2} \frac{v_F^4}{T^3} }\right] ,\label{eq:al2}
\end{eqnarray}
\begin{eqnarray}
&& \bar{\sigma}^{a)}_C(\bm{ q}) = C \cdot \frac{\pi^2 u_{(2,1)}(t,\gamma)}{192} \frac{v_F^3}{k_F T^2} N(0)^2  W_0 \tau^2 q^2_z, \nonumber \\
&& \\
&& \bar{\sigma}^{b)}_C(\bm{ q}) = C \cdot \frac{ w_{(2,1)}(t,\gamma)}{96} \frac{v_F^3}{k_F T^2} N(0)^2  W_0 \tau^2 q^2_z, \nonumber \\
&&\label{eq:sig2}
\end{eqnarray}
where $t=2\pi T/ \Gamma$ and $\gamma = 1/2\pi \tau T$, and the definitions of dimensionless functions $u_{(i,j)} (t,\gamma)$ and $w_{(i,j)} (t,\gamma) $ are given in Sec. \ref{dimensionless}.
In clean limit ($\gamma \to 0$),
\begin{eqnarray}
&& \bar{\alpha}^{a)}_C(\bm{ q}) \nonumber \\
&& = C \cdot N(0)^2  W_0 \tau^2 q^2_z \times \nonumber \\
&& \quad \left[ \frac{5 \pi^2 u_{(1,0)}(\frac{2\pi T}{\Gamma}, 0)  }{96} \frac{v_F^2}{k_F^2 T} - \frac{ \pi^2 u_{(2,1)}(\frac{2\pi T}{\Gamma}, 0)  }{1920 } \frac{v_F^4}{T^3}  \right] ,
\nonumber \\
&& \label{ti_alpha_a}\\
&& \bar{\alpha}^{b)}_C(\bm{ q})  \nonumber \\
&& =  C \cdot N(0)^2  W_0 \tau^2 q^2_z \times \nonumber \\
&& \quad  \left[ - \frac{5 \pi^2 u_{(1,0)}(\frac{2\pi T}{\Gamma}, 0) }{96} \frac{v_F^2}{k_F^2 T} - \frac{  w_{(2,1)}(\frac{2\pi T}{\Gamma}, 0)  }{960} \frac{v_F^4}{T^3} \right], \nonumber \\
&&  \label{ti_alpha_b}\\
&& \bar{\sigma}^{a)}_C(\bm{ q}) \nonumber \\
&& = C\cdot \frac{\pi^2 u_{(2,1)}(\frac{2\pi T}{\Gamma}, 0)}{192} \frac{v_F^3}{k_F T^2} N(0)^2  W_0 \tau^2 q^2_z, \label{ti_sigma_a} \\
&& \bar{\sigma}^{b)}_C(\bm{ q}) \nonumber \\
&& = C \cdot \frac{ w_{(2,1)}(\frac{2\pi T}{\Gamma}, 0)}{96} \frac{v_F^3}{k_F T^2} N(0)^2  W_0 \tau^2 q^2_z. \label{ti_sigma_b}
\end{eqnarray}

Finally, we complete the calculations of the Nernst and Hall conductivities,
which are given by the integral over ${\bm q}$:
\begin{eqnarray}
&& \alpha^{Kubo}_{xy} = \sum_{i=a,b,c} (-2 e T) \sum_{\bm{ q}\, C=\pm 1}  \tilde{L}'_{C}(\bm{ q}, {\omega_q=0};H)  \bar{\alpha}^{i)}_C(\bm{ q}), \nonumber \\
&& \label{alphaKuboLalpha} \\
&& \sigma_{xy} 
 =   \sum_{i=a,b,c} (2 e^2 T ) \sum_{\bm{ q} \, C= \pm 1}  \tilde{L}'_{C}(\bm{ q}, {\omega_q=0};H)  \bar{\sigma}^{i)}_C(\bm{ q}), \nonumber \\
&& \label{sigmaLsigma}
\end{eqnarray}
where the chirality-dependent part of the BPS is given by
\begin{eqnarray}
\tilde{L}' _C({\bm q},\omega_q;H) = C( 5eH/4k_F^2g) \left[ L({\bm q},\omega_q)\right]^2.
\end{eqnarray}
Using the expressions (\ref{ti_alpha_a}-\ref{ti_sigma_b}) and (\ref{L_momentum}),
we encounter ultraviolet divergence in the calculation of (\ref{alphaKuboLalpha}) and (\ref{sigmaLsigma}).
Then, we introduce the cutoff momentum $\Lambda$.
It is appropriate to set $\Lambda$ as the same order as $1/\xi$, where $\xi$ is the coherence length. The precise definition of
$\xi$ in this letter is given by Eq.(\ref{eq:coh}) shown later.
The reason is that the expansion of $L^{-1}$ up to the second order of ${\bm q}$, (\ref{L_momentum}), is justified
when $|{\bm q}|$ is sufficiently smaller than $1/\xi$,
and, then, $\tilde{L}'_C$ rapidly decreases as $|{\bm q}|$ increases in the region $|{\bm q}| > 1/\xi $, because of higher-order terms, as discussed in Sec. \ref{FPunderzeromag}.
Besides, in $|{\bm q}| < \Lambda$,
the expressions (\ref{ti_alpha_a}-\ref{ti_sigma_b}) are also justified,
since $\xi k_F \gg 1$ is satisfied in almost all superconductors (indeed, $\xi k_F \sim 50$ in URu$_2$Si$_2$ \cite{Schlabitz}).
To simplify the expressions of the final results,
we assume that the domain of the integral is anisotropic: $\sum_i a_i q_i^2 \leq (a_x^2 a_z^3)^{1/5} \Lambda^2$,
where the numerical factors are given by $a_x=a_y = 6/7$ and $a_z=9/7$.
Then, we obtain,
\begin{eqnarray}
&& \alpha^{Kubo}_{xy \, chiral} \nonumber \\
&& =  \frac{f\left( \frac{2\pi T}{\Gamma} \right)}{2304}
\frac{ \tau^2 e^2 W_0 v^4_F \Lambda H }{\xi^4 g k_F^2 T^2}  \nonumber \\
&& \quad \times  \left(1-\frac{3  \sqrt{ \varepsilon} \arctan \left( \frac{\xi \Lambda}{\sqrt{\varepsilon}} \right) }{2 \xi\Lambda } + \frac{1}{2+ 2 (\xi \Lambda)^2 / \varepsilon } \right), \nonumber \\
&& \\
&& \sigma_{xy \, chiral} \nonumber \\
&& = \frac{5 f\left( \frac{2\pi T}{\Gamma} \right)}{1152}\frac{\tau^2 e^3  W_0 v_F^3  \Lambda }{\xi^4 g k^3_F T}  \nonumber \\
&&\quad \times   \left(1-\frac{3  \sqrt{ \varepsilon} \arctan \left( \frac{\xi \Lambda}{\sqrt{\varepsilon}} \right) }{2 \xi\Lambda } +\frac{1}{2+ 2 (\xi \Lambda)^2 / \varepsilon } \right), \nonumber \\
&&
\end{eqnarray}
in clean limit,
where $\xi'=(a^2_x a^3_z)^{1/10} \approx 1.05 \xi$ and 
$f(t)= u_{(2,1)}(t,0) +(2/\pi^2) w_{(2,1)}(t, 0)$.
Since $\xi' \sim \xi$, we use $\xi$ instead of $\xi'$ in the following.
For $T\sim T_c$, the above equations are reduced to Eqs. (4) and (5) in the main text.

\subsection{Momentum Representation of Fluctuation Propagators under Zero Magnetic Field}
\label{FPunderzeromag}
In this section, we derive the expression of the fluctuation propagator of $d_{zx} \pm {\rm i} d_{zy}$-wave superconductors under zero magnetic field,
which is used in Sec. \ref{Derivations_of_Nernst_and_Hall}.
It is given by,
\begin{eqnarray}
&&L^{-1}({\bm q},\omega_q) =-1/g +\Pi({\bm q},\omega_q),  \\
&&\Pi ({\bm q},\omega_q) = T \sum_{n} \Pi_{\varepsilon_n} ({\bm q},\omega_q), \\
&&\Pi_{\varepsilon_n} ({\bm q},\omega_q) =\nonumber \\
&& \sum_{{\bm k}} | \phi \left( {\bm k}  \right) |^2 G\left({\bm k} +\frac{{\bm q}}{2} , \varepsilon_{n+q} \right) G\left(-{\bm k} +\frac{{\bm q}}{2} , -\varepsilon_{n} \right), \nonumber \\
&&
\end{eqnarray}
where $G({\bm k}, \varepsilon_{k})^{-1}={\rm i}\tilde{\varepsilon}_k -\xi_{{\bm k}}={\rm i} ( \varepsilon_{k} + {\rm sgn  (\varepsilon_{k} )/2\tau}) -\xi_{{\bm k}} $,
$\phi(\bm{ k})= \sqrt{15/2} k_z (k_x + {\rm i} k_y)/k^2_{F}$.
Now, replacing the sum with the integral over the energy and average over the Fermi surface, $<\cdots >_{\hat{{\bm k}}}$,
we obtain
\begin{eqnarray}
 && \Pi_{\varepsilon_n} ({\bm q},\omega_q) = \nonumber \\
 &&  2 \pi N(0) \theta(\varepsilon_{n+q} \varepsilon_n) 
\left< \frac{ \left| \phi ({\bm k} ) \right|^2 } { | \tilde{ \varepsilon}_{n+q} + \tilde{\varepsilon}_{n}| + {\rm i} \Delta \xi ({\bm k},{\bm q}) } \right> _{\hat{{\bm k}}},\nonumber \\
\end{eqnarray}
where $N(0)$ is the density-of-state at the Fermi surface,
$\theta$ is the Heaviside step function,
 and $\Delta \xi ({\bm k},{\bm q})  = \xi_{{\bm k} +{\bm q}/2} - \xi_{{\bm k} - {\bm q}/2} $.
Expanding it with respect to ${\bm q}$,
we obtain the expression up to the quadratic term:
\begin{eqnarray}
&& \Pi_{\varepsilon_n} ({\bm q},\omega_q) \nonumber \\
&&= \frac{2\pi N(0) \theta(\varepsilon_{n+q} \varepsilon_n)}{ | \tilde{ \varepsilon}_{n+q} + \tilde{\varepsilon}_{n}| } \times   \nonumber \\
 &&\left[ < | \phi \left( {\bm k}  \right) |^2 >_{\hat{{\bm k}}} - \frac{ < | \phi \left( {\bm k}  \right)|^2 ( {\bm v}_{{\bm k}} \cdot {\bm q} )^2  >_{\hat{{\bm k}}}}{ | \tilde{ \varepsilon}_{n+q} + \tilde{\varepsilon}_{n}|^2 } \right] \nonumber \\
&&= \frac{2\pi N(0) \theta(\varepsilon_{n+q} \varepsilon_n)}{ | \tilde{ \varepsilon}_{n+q} + \tilde{\varepsilon}_{n}| } 
\left[  1- \frac{ v^2_F \sum_{i=x,y,z}a_i q^2_i  }{ | \tilde{ \varepsilon}_{n+q} + \tilde{\varepsilon}_{n}|^2 } \right], \nonumber \\
\end{eqnarray}
where $v_F$ is the Fermi velocity and $a_{x}=a_{y}=6/7$ and $a_z=9/7$ are numerical factors which reflect anisotropy of $V^{\pm } (\bm{k},\bm{k}')$.
Here, this expansion is justified when 
$\Delta \xi ({\bm k},{\bm q})$ is sufficiently smaller than $| \tilde{ \varepsilon}_{n+q} + \tilde{\varepsilon}_{n}| $.
Due to the factor $\theta(\varepsilon_{n+q} \varepsilon_n)$, 
$| \tilde{ \varepsilon}_{n+q} + \tilde{\varepsilon}_{n}| $ is equal to or larger than $2 \pi T$,
and, therefore, this condition is read as $v_F q \ll T \iff q \ll 1/\xi$, where
the coherence length $\xi$ used in this letter is precisely defined 
by Eq.(\ref{eq:coh}).

Now, we take the sum over $n$,
in which we introduce the cutoff energy $\omega_D$ for the pairing interaction and the upper limit of the frequency sum $N_{max}= \omega_D / 2\pi T$ to remove the ultraviolet logarithmic divergence of the first sum:
\begin{eqnarray}
&&\frac{ \Pi ({\bm q},\omega_q)}{N(0)} 
= \psi \left( \frac{1}{2} + \frac{|\omega_q|}{4 \pi T} + \frac{\omega_D}{2 \pi T} + \frac{1}{4 \pi T \tau} \right) \nonumber \\
&& \qquad \qquad \quad -\psi \left( \frac{1}{2}  + \frac{|\omega_q|}{4 \pi T}  + \frac{1}{4 \pi T \tau} \right) \nonumber \\
&& \qquad \qquad \quad + \frac{v^2_F a_i q^2_i }{2 (4 \pi T)^2} \psi'' \left( \frac{1}{2}  + \frac{1}{4 \pi T \tau}  \right).
\end{eqnarray}
The superconducting transition temperature $T_c$ is defined by $L^{-1} ({\bm 0},0)|_{T=T_c} = 0$,
which is rewritten into
\begin{eqnarray}
\psi \left( \frac{1}{2} + \frac{\omega_D}{2 \pi T} + \frac{1}{4 \pi T_c \tau} \right) 
 -\psi \left( \frac{1}{2}  + \frac{1}{4 \pi T_c \tau} \right) = \frac{1}{g N(0)}. \nonumber \\
 \label{def_of_Tc}
\end{eqnarray}
Therefore, we obtain the expression for the fluctuation propagator
in the vicinity of $T_c$, i.e. $\varepsilon = \ln T/T_c \ll 1$:
\begin{eqnarray}
&& L^{-1}({\bm q},\omega_q)  \nonumber \\
&&= - N(0) \left[ \varepsilon + \psi \left( \frac{1}{2} +\frac{|\omega_q|}{4 \pi T}+  \frac{1}{4 \pi T \tau} \right) \right.  \nonumber \\
&& \quad \left. -\psi \left( \frac{1}{2} + \frac{1}{4 \pi T \tau} \right)+ \sum_i \xi^2_i( \tau ) q^2_i    \right], \label{L_momentum}
\end{eqnarray}
where the coherence length is given by $\xi^2_{i} (\tau) = a_i \xi^2(\tau) $, where $ \xi^2(\tau) = - v^2_{F} \psi''(1/2+1/4\pi T \tau)  /6(4 \pi T)^2 $.
The coherence length used in this letter is defined by
\begin{eqnarray}
\xi=\xi(\tau = \infty).
\label{eq:coh}
\end{eqnarray}
Eq.(\ref{L_momentum}) is the main result of this section.


\subsection{Dimensionless Functions}
\label{dimensionless}
In this section, we give the definitions of the dimensionless functions which appear in the formulae of the Nernst and Hall conductivities presented in
the previous section.
We also obtain their approximated but explicit expressions.
The main result of this section is that the dimensionless function $f(t)$ which appears in Eqs.(4) and (5) in the main text is
well approximated by a smooth function $\bar{f}^{app}(t)$ (\ref{eq:f_app_bar})
for temperature regions where superconducting fluctuations are strong.
We use it for the numerical calculation of temperature dependences of transport coefficients shown in FIG. 3 in the main text.

\subsubsection{{\rm Definitions}}
The definitions of dimensionless functions which appear in Eqs.(\ref{eq:al1})-(\ref{eq:sig2}) are 
\begin{eqnarray}
&&u_{(i,j)} (t,\gamma)= \frac{(-1)^{i+j+1}(i+j-1)!}{\psi^{(i+j)}(\frac{1}{2})}
 \left[  \frac{2\pi T}{\omega_l} \times \right. \nonumber \\
&& \left. \sum_{0 \leq n,m  \leq l-1} \mathsmaller{ \frac{1}{(n+\frac{1}{2}+\frac{\gamma}{2})^i(m+\frac{1}{2}+\frac{\gamma}{2})^j(1+t |n-m|)}}  \right]_{\substack{ \mathsmaller{ {\rm i} \omega_l \to \omega + {\rm i} 0 ,} \\ \omega \to 0}},  \nonumber \\
\label{uij}\\
&& w_{(i,j)} (t,\gamma) = \frac{(i-1)! (j-1)!}{(i+j-2)! \pi^2}  \left[  \frac{2\pi T}{\omega_l} \times \right. \nonumber \\
&& \left. \sum_{0 \leq n,m  \leq l-1}  \mathsmaller{ \frac{1}{(n+\frac{1}{2}+\frac{\gamma}{2})^i(l-m-\frac{1}{2}+\frac{\gamma}{2})^j(1+t |n-m|)} } \right]_{ \substack{ {\rm i} \omega_l \to \omega + {\rm i} 0 , \\ \omega \to 0}},  \nonumber \\
\label{wij}
\end{eqnarray}
where $(i,j)$ are nonnegative integers and the domains of definitions are $i+j \geq 1$  for $u_{(i,j)} (t)$, and $i,j \geq 1 $ for $w_{(i,j)} (t,\gamma)$.
An important property of these functions is that at $t=0$,
\begin{eqnarray}
u_{(i,j)} (t=0, \gamma)=w_{(i,j)} (t=0,\gamma) = 0. \label{t0value}
\end{eqnarray}
Also, their normalization factors are determined by the following conditions:
\begin{eqnarray}
&& u_{(i,j)} (t=\infty, \gamma=0)= 1, \label{norm1}\\
&& w_{(i,j)} (t=\infty,\gamma) = \frac{1}{\gamma ^{i+j-1}} +\mathcal{O}\left(\frac{1}{\gamma ^{i+j-2}} \right) \quad {\rm as} \, \,  \gamma \to 0. \nonumber \\
\label{norm2}
\end{eqnarray}
Eqs. (\ref{t0value}-\ref{norm2}) are proved in Sec. \ref{proof_of_relations}.

The dimensionless function $f(t)$ which appears in Eqs.(4) and (5) in the main text, and Eqs.(S.54) and (S.55) in  Supplemental Material is defined by using $u_{(i,j)} (t,\gamma)$ and $w_{(i,j)} (t,\gamma)$ as,
$$ f(t)= u_{(2,1)}(t,0) +(2/\pi^2) w_{(2,1)}(t, 0)$$.

\subsubsection{{\rm Approximation Functions}}
We, here, introduce analytically-solvable approximation functions for the dimensionless functions (\ref{uij}) and (\ref{wij}).
They are defined by the analytic continuation of functions of the Matsubara frequencies:
\begin{eqnarray}
&&u^{app}(t) = \left.  \frac{2\pi T}{\omega_l} \sum_{0 \leq n,m  \leq l-1} \frac{1}{1+t |n-m|}  \right|_{ \substack{  \mathsmaller{ {\rm i} \omega_l \to \omega + {\rm i} 0, } \\   \mathsmaller{ \omega \to 0} } } , \nonumber \\
\label{u_app_def} \\
&& w^{app}_{i+j}(t,\gamma) \nonumber \\
&& = \left. \frac{2\pi T}{\omega_l} \sum_{0 \leq n,m  \leq l-1}  \mathsmaller{ \frac{1}{(n-m+l+\gamma)^{i+j-1}} \frac{1}{1+t |n-m|}  } \right|_{ \substack{  \mathsmaller{ {\rm i} \omega_l \to \omega + {\rm i} 0, } \\   \mathsmaller{ \omega \to 0} } }, \nonumber \\
 \label{w_app_def}
\end{eqnarray}
and they can be rewritten into compact expressions:
\begin{eqnarray}
&& u^{app}(t) = -1 - \frac{2}{t} + \frac{2}{t^2}  \psi' \left( \frac{1}{t} \right), \label{u_app} \\
&& w^{app}_{i+j}(t,\gamma) = \frac{(-1)^{i+j-2} }{(i+j-2)!} \frac{\partial^{i+j-2} }{\partial \gamma ^{i+j-2}} w^{app}_{2} (t,\gamma), \nonumber \\
&& w^{app}_{2}(t,\gamma) = \frac{1}{(1/t - \gamma ) t} (\gamma \psi'(\gamma) -\frac{1}{t} \psi'(1/t) ) \nonumber \\
&& \qquad \qquad - \frac{1}{(1/t + \gamma ) t} (\gamma \psi'(1+\gamma) -\frac{1}{t} \psi'(1/t) ) -\frac{1}{\gamma}, \nonumber \\
\label{w_app}
\end{eqnarray}
which are derived in Sec. \ref{proof_of_relations}.

We can expect that 
$u^{app}(t)$ and $w^{app}_{i+j}(t ,\gamma)$ with small $\gamma$ are good approximation functions for $u_{(i,j)}(t,\gamma=0)$ and $w_{(i,j)}(t,\gamma)$, respectively, because of the following reason.
They have the same asymptotic behaviors as the original functions
for $t \to \infty$, and 
also take
the same values as the original ones at $t=0$:
\begin{eqnarray}
\left.  \frac{\partial^n u_{(i,j)}(t,\gamma=0)}{\partial t^n}    \right|_{t = \infty} &=&\displaystyle{ \left.  \frac{\partial^n   u^{app}(t) }{\partial t^n}  \right|_{t = \infty}} , \nonumber \\
&=& \delta_{n,0} \quad  {\rm for} \, n \geq 0,  \nonumber \\
&& \label{u_uapp_tinf}  \\
u_{(i,j)}(t=0,\gamma) &=& u^{app}(t=0)  \nonumber \\
&=&0, \label{u_uapp_t0}
\end{eqnarray}
and
\begin{eqnarray}
w_{(i,j)}(t= \infty ,\gamma)  &=& w^{app}_{i+j}(t= \infty ,\gamma)\nonumber \\
&=& \frac{1}{\gamma ^{i+j-1}} +\mathcal{O}\left(\frac{1}{\gamma ^{i+j-2}} \right) \quad {\rm as} \, \,  \gamma \to 0, \nonumber \\
\label{w_uapp_tinf_0}
\end{eqnarray}
\begin{eqnarray}
\left.  \frac{\partial^n w_{(i,j)}(t,\gamma)}{\partial t^n}    \right|_{\gamma:{\rm fixed}, t = \infty} &=&\displaystyle{ \left.  \frac{\partial^n   w^{app}_{i+j}(t,\gamma) }{\partial t^n}  \right|_{\gamma:{\rm fixed},t = \infty}} \nonumber \\
&=&0,   \quad  {\rm for} \, n \geq 1,  \nonumber \\
&&  \label{w_wapp_tinf_non0} \\
w_{(i,j)}(t=0,\gamma) &=& w^{app}_{i+j}(t=0,\gamma) =0. \label{w_wapp_t0}
\end{eqnarray}
We will prove these relations in Sec.\ref{proof_of_relations}.

\begin{figure}
 \includegraphics[clip, width=40mm]{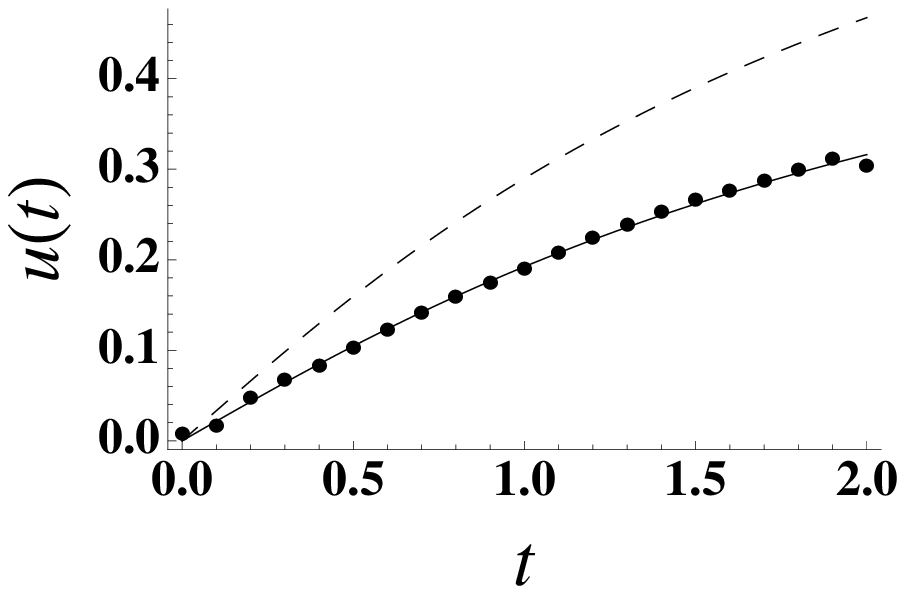}~ 
 \includegraphics[clip, width=40mm]{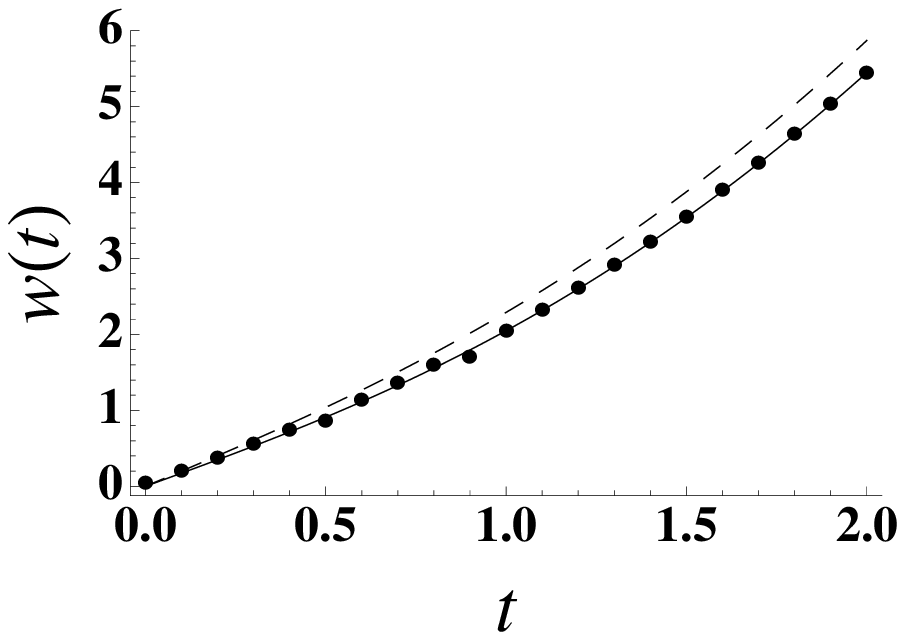}\\ 
 \includegraphics[clip, width=70mm]{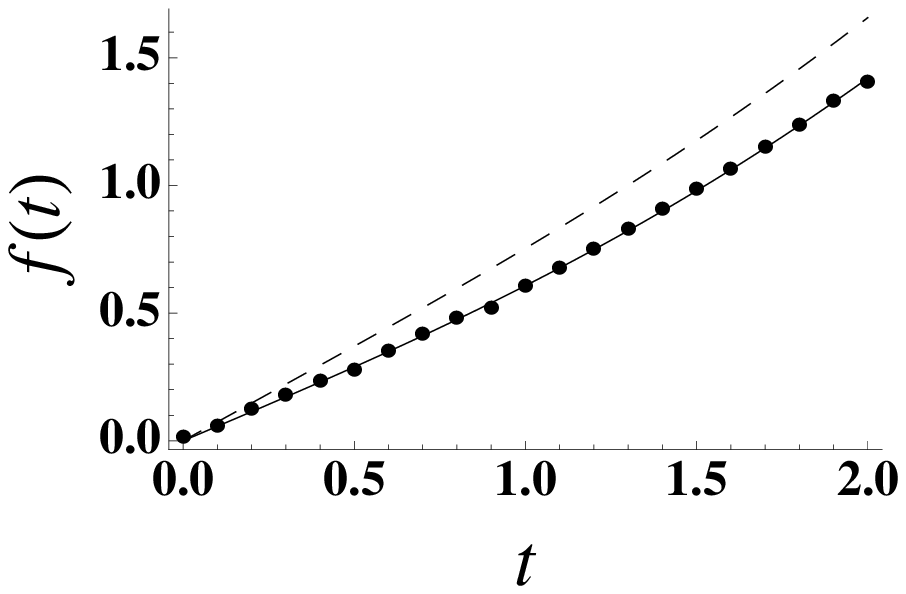} 
 \centering
 \caption{{\it Upper Left} ({\it Upper Right, Lower}): Numerical result for $u_{(2,1)}(t,0)$ ($w_{(2,1)}(t,0)$, $f(t)$) (dots),
    approximation function $u^{app}(t) $ ($w^{app}_{3}(t,0)$, $f^{app}(t)$) (dashed line),
and  modified approximation function $\bar{u}^{app}(t) $ ($\bar{w}^{app}_{3}(t,0)$, $\bar{f}^{app}(t)$) (solid line). 
The first and third ones coincide with each other quite well.}          
 \label{uwf}
 \end{figure}

In FIG. \ref{uwf}, we plot these approximation functions (dashed line) and numerical estimations of Eqs. (\ref{uij}) and (\ref{wij}) (dots) which are obtained by the Pad\'e method.
In this plot, we focus on $u_{(2,1)}(t,0)$, $w_{(2,1)}(t,0)$, and their sum, $f(t)=u_{(2,1)}(t,0)+ (2/\pi^2)w_{(2,1)}(t,0)$,
which appears in Eqs. (4) and (5).
Here, the range of the plot is set to be $0 < t < 2.0$,
which covers the region of superconducting fluctuations $t_c < t < x t_c$, $x \sim 3$,
with $t_c = 2 \pi T_c / \Gamma =0.5$.
In this calculations, 
we used the material parameters of URu$_2$Si$_2$, i.e. $T_c \sim 1.5 {\rm K}$ and $\Gamma \sim 1.5 {\rm meV}$ \cite{KasaharaPRL, Bourdarot}.
As seen from FIG. \ref{uwf}, the $t$-dependences of the approximation functions are qualitatively similar to the original functions.
However, there are slight quantitative differences.
Then, to improve the approximation functions, 
we scale them as 
\begin{eqnarray}
&& \bar{u}^{app}(t) = c_u \cdot u^{app}(c'_u \cdot t), \\
&& \bar{w}^{app}_{3}(t,0) = c_w \cdot w^{app}_{3}(c'_w \cdot t,0), \\
&& \bar{f}^{app}(t)= \bar{u}^{app}(t)+ \frac{2}{\pi^2} \bar{w}^{app}_{3}(t,0) 
 \label{eq:f_app_bar},
\end{eqnarray}
where the scaling parameter constants are obtained by fitting the numerical data, and we find $c_u= 0.72$, $c'_u= 0.91$, $c_w= 0.76$, and $c'_w= 1.14$.
The improved approximation functions are also shown in FIG. \ref{uwf} (solid line) and we see that they coincide with 
the original functions quite well.
Therefore, we use these smooth functions to calculate temperature dependences of transport coefficients in the main text.

\subsubsection{{\rm Proofs of Relations}}
\label{proof_of_relations}
In this subsection, we give proofs of relations used in the previous sections.

\paragraph{{\rm Proofs of Eq. (\ref{t0value})}}
\indent

We can easily verify the following relation,
\begin{eqnarray}
&&u_{(i,j)} (t=0,\gamma) \nonumber \\
&& = c_{ij} \left. \mathsmaller{ \frac{2\pi T}{\omega_l}} \sum_{0 \leq n,m  \leq l-1} \mathsmaller{ \frac{1}{(n+\frac{1}{2}+\frac{\gamma}{2})^i(m+\frac{1}{2}+\frac{\gamma}{2})^j}}  \right|_{\substack{ \mathsmaller{ {\rm i} \omega_l \to \omega + {\rm i} 0 ,} \\ \omega \to 0}}  \nonumber \\
&&=  c_{ij} \left[  \frac{2\pi T}{\omega_l} \frac{\psi^{(i-1)}(\frac{1}{2})-\psi^{(i-1)}(\frac{1}{2} + \frac{\omega_l}{2\pi T})}{(-1)^{i}(i-1)!}  \right. \nonumber \\
&&\quad  \times \left. \frac{\psi^{(i-1)}(\frac{1}{2})-\psi^{(i-1)}(\frac{1}{2} + \frac{\omega_l}{2\pi T})}{(-1)^{i}(i-1)!}  \right]_{ \substack{  \mathsmaller{ {\rm i} \omega_l \to \omega + {\rm i} 0, } \\   \mathsmaller{ \omega \to 0} } } \nonumber \\
&& =0 \label{wij2}
\end{eqnarray}
where $c_{ij}=(-1)^{i+j+1}(i+j-1)!/\psi^{(i+j)}(\frac{1}{2})$.
Similar calculation leads to $w_{(i,j)} (t,\gamma) = 0$.
Then Eqs. (\ref{t0value}) is proved.

 \phantom{*}

\paragraph{{\rm Proofs of Eqs. (\ref{norm1}) and (\ref{norm2})}}
\indent

First, we prove Eq. (\ref{norm1}).
Owing to the factor $1/(1+t|n-m|)$, only terms satisfying $n=m$ contribute to the sum at $t \to \infty$. 
Therefore, 
\begin{eqnarray}
&&u_{(i,j)} (t=\infty,\gamma=0) \nonumber \\
&&= c_{ij}   \left. \frac{2\pi T}{\omega_l} \sum_{n=0,1, \cdots,  l-1} \frac{1}{(n+\frac{1}{2})^{i+j}}  \right|_{  \substack{   \mathsmaller{ {\rm i} \omega_l \to \omega + {\rm i} 0 ,} \\  \mathsmaller{ \omega \to 0} } } \nonumber \\
&&=\scriptstyle{ c_{ij}  \left.  \frac{2\pi T}{\omega_l} \frac{\psi^{(i+j-1)}(\frac{1}{2})-\psi^{(i+j-1)}(\frac{1}{2} + \frac{\omega_l}{2\pi T})}{(-1)^{i+j}(i+j-1)!}  \right|_{ \substack{  \mathsmaller{ {\rm i} \omega_l \to \omega + {\rm i} 0, } \\   \mathsmaller{ \omega \to 0} } } }\nonumber \\
&& = c_{ij}/  c_{ij} =1.
\end{eqnarray}

Next, we prove Eq. (\ref{norm2}).
Taking the limit $t\rightarrow \infty$, we have,
\begin{eqnarray}
&& w_{(i,j)} (t=\infty,\gamma) = \frac{(i-1)! (j-1)!}{(i+j-2)! \pi^2}   \nonumber \\
&&\times \scriptstyle{ \left. \frac{2\pi T}{\omega_l} {\displaystyle \sum_{n=0,1, \cdots,  l-1}} \frac{1}{(n+\frac{1}{2}+\frac{\gamma}{2})^i(l-n-\frac{1}{2}+\frac{\gamma}{2})^j}  \right|_{ \substack{  \mathsmaller{ {\rm i} \omega_l \to \omega + {\rm i} 0, } \\   \mathsmaller{ \omega \to 0} } } }. \nonumber\\
\end{eqnarray}
Now, let 
\begin{eqnarray}
&& \tilde{w}_{(i,j)} (\gamma) \nonumber \\
&& = \left. \scriptstyle{ \frac{2\pi T}{\omega_l}} {\displaystyle \sum_{n=0,1, \cdots,  l-1}}
\scriptstyle{ \frac{1}{(n+\frac{1}{2}+\frac{\gamma}{2})^i(l-n-\frac{1}{2}+\frac{\gamma}{2})^j}}  \right|_{ \substack{  \mathsmaller{ {\rm i} \omega_l \to \omega + {\rm i} 0, } \\   \mathsmaller{ \omega \to 0} } } . \nonumber \\
\end{eqnarray}
Then, 
\begin{eqnarray}
&& \tilde{w}_{(i,j)} (\gamma) \nonumber \\
&& \frac{1}{\gamma } ( \tilde{w}_{(i,j-1)} (\gamma)+\tilde{w}_{(i-1,j)}(\gamma) ) \nonumber \\
&& = \frac{1}{\gamma^2}  ( \tilde{w}_{(i,j-2)} (\gamma)+2\tilde{w}_{(i-1,j-1)}(\gamma)+\tilde{w}_{(i-2,j)}(\gamma) ) \nonumber \\
&& \vdots \nonumber \\
&& = \frac{(i+j-2)!}{(i-1)! (j-1)!} \frac{1}{\gamma ^{i+j-1}} ( \tilde{w}_{(1,0)}(\gamma) +\tilde{w}_{(0,1)}(\gamma) ) \nonumber \\
&& \quad + c'_1  \frac{1}{\gamma ^{i+j-2}} ( \tilde{w}_{(2,0)}(\gamma) +\tilde{w}_{(0,2)} (\gamma)) \nonumber \\
&& \quad + c'_2  \frac{1}{\gamma ^{i+j-3}} ( \tilde{w}_{(3,0)}(\gamma) +\tilde{w}_{(0,3)} (\gamma)) \nonumber \\
&& \quad + \cdots
\end{eqnarray}
where $c'_1, c'_2  \cdots$ are constants. By using 
\begin{eqnarray}
&& \tilde{w}_{(i,0)} (\gamma=0) \nonumber \\ 
&&= \left.  \frac{2\pi T}{\omega_l} \sum_{n=0,1, \cdots,  l-1} \frac{1}{(n+\frac{1}{2})^i}   \right|_{ \substack{  \mathsmaller{ {\rm i} \omega_l \to \omega + {\rm i} 0, } \\   \mathsmaller{ \omega \to 0} } } \nonumber \\
&&=  \frac{(-1)^{i+1}\psi^{(i)}(\frac{1}{2})} {(i-1)!},
\end{eqnarray}
we obtain
\begin{eqnarray}
&& w_{(i,j)} (t=\infty,\gamma) \nonumber \\
&& =  \frac{(i-1)! (j-1)!}{(i+j-2)! \pi^2} \tilde{w}_{(i,j)} (\gamma) \nonumber \\
&& = \frac{1}{\gamma ^{i+j-1}} +\mathcal{O}\left(\frac{1}{\gamma ^{i+j-2}} \right) \qquad {\rm as} \quad \gamma \to 0. \nonumber \\
\end{eqnarray}
Then, the normalization conditions (\ref{norm1}) and (\ref{norm2}) are proved.

 \phantom{*}

\paragraph{{\rm Proofs of Eqs. (\ref{u_app}) and (\ref{w_app})}}
\indent

Now, introducing $N=n-m$ and dividing the region of summation in (\ref{u_app_def}) into $\displaystyle \sum_{n\leq m} + \sum_{n \geq m} - \sum_{n=m}$,
we obtain,
\begin{eqnarray}
u^{app}(t) &=& 2 \times  \left. \left[ \frac{2\pi T}{\omega_l} \sum_{N=0,1, \dots ,l-1} \frac{l-N}{1+tN} \right]  \right|_{ \substack{  \mathsmaller{ {\rm i} \omega_l \to \omega + {\rm i} 0, } \\   \mathsmaller{ \omega \to 0} } } - 1 \nonumber \\
&=& -1 -\frac{2}{t} +\frac{2}{t^2} \left. \left[ \mathsmaller{  \frac{2\pi T}{\omega_l}} \sum_{N=0,1, \dots ,l-1}  \mathsmaller{ \frac{1}{N+1/t}} \right]  \right|_{ \substack{  \mathsmaller{ {\rm i} \omega_l \to \omega + {\rm i} 0, } \\   \mathsmaller{ \omega \to 0} } }\nonumber \\
&& + \frac{2}{t}  \left. \left[ \sum_{N=0,1, \dots ,l-1} \frac{1}{1+tN} \right]  \right|_{ \substack{  \mathsmaller{ {\rm i} \omega_l \to \omega + {\rm i} 0, } \\   \mathsmaller{ \omega \to 0} } }  \nonumber \\
&=&  -1 - \frac{2}{t} + \frac{2}{t^2}  \psi' \left( \frac{1}{t} \right).
\end{eqnarray}
From it we can derive Eq. (\ref{u_app}).
Furthermore, Eq. (\ref{w_app}) can be obtain by similar calculations.

 \phantom{*}

\paragraph{{\rm Proofs of Eqs. (\ref{u_uapp_tinf}) and (\ref{u_uapp_t0})}}
\indent

First, we prove Eq. (\ref{u_uapp_tinf}).
The $n=0$ case  immediately follows from Eq. (\ref{norm1}) and
the explicit expression for $u_{app}(t)$, (\ref{u_app}).
For $n \geq 1$,
\begin{eqnarray}
&&\left.  \frac{\partial^n u_{(i,j)}(t,\gamma=0)}{\partial t^n}    \right|_{t = \infty} \nonumber \\
&&=c_{ij} \left. \mathsmaller{ \frac{2\pi T}{\omega_l}} \sum_{0 \leq a,b  \leq l-1}d_{ij}^{ab} (-1)^n n!
\mathsmaller{ \frac{|a-b|^n}{(1+t |a-b|)^{n+1}}}  \right|_{\substack{ \mathsmaller{ {\rm i} \omega_l \to \omega + {\rm i} 0 ,} \\ \omega \to 0}}, \label{u_n-dif_t} \nonumber\\
\end{eqnarray}
where $d_{ij}^{ab}= 1/(a+\frac{1}{2})^i(b+\frac{1}{2})^j $.
Now, owing to the factor $1/(1+ t |a-b|)^{n+1}$,
the contributions from the terms satisfying $a \neq b$ become zero in $t \to \infty$ limit.
Moreover, the other terms, which satisfy $a = b$, are also zero for $n \geq 1$
on account of the factor $|a-b|^n$.
Therefore, we find that Eq. (\ref{u_n-dif_t}) is zero in $t \to \infty$ limit.
A similar calculation leads $\left. \partial^n   u^{app}(t) / \partial t^n  \right|_{t = \infty} = 0$,
which also directly follows from the explicit expression (\ref{u_app}).
Then, Eq. (\ref{u_uapp_tinf}) for any $n \geq 0$ is proven.

On the other hand, Eq. (\ref{u_uapp_t0}) instantly follows from Eq. (\ref{t0value}) and
a calculation similar to Eq. (\ref{wij2}).

 \phantom{*}

\paragraph{{\rm Proofs of Eqs. (\ref{w_uapp_tinf_0}), (\ref{w_wapp_tinf_non0}), and (\ref{w_wapp_t0})}}
\indent

The equivalence between the LHS and RHS of Eq. (\ref{w_uapp_tinf_0}) has been already proven (\ref{norm2}).
Then, we now prove the equivalence between the middle one and the RHS.
Owing to the factor $1/(1+t|n-m|)$, only terms satisfying $n=m$ contribute to the sum at $t \to \infty$:
\begin{eqnarray}
&& w^{app}_{i+j}(t= \infty ,\gamma) \nonumber \\
&&=  \left. \frac{2\pi T}{\omega_l} \sum_{ n= 0, 1, \cdots , l-1}  \mathsmaller{ \frac{1}{(l+\gamma)^{i+j-1}}  } \right|_{ \substack{  \mathsmaller{ {\rm i} \omega_l \to \omega + {\rm i} 0, } \\   \mathsmaller{ \omega \to 0} } } \nonumber \\
&&= \frac{1}{\gamma^{i+j-1}}.
\end{eqnarray}
Then, the whole of Eq. (\ref{w_uapp_tinf_0}) is proven.

On the other hand, Eqs. (\ref{w_wapp_tinf_non0}) and (\ref{w_wapp_t0})
can be derived with techniques similar to those used in the derivations of Eq (\ref{u_uapp_tinf}) for $n \geq 1$ and Eq. (\ref{u_uapp_t0}).

\subsection{Magnetization}
In this section we discuss the magnetization and derive Eqs. (7) and (8).
In this letter $\hat{A}$ means an matrix whose element is $A({\bm x},{\bm y})$ and $\hat{1}_{{\bm x},{\bm y}} = \delta ({\bm x}-{\bm y})$.
The free energy in the presence of the magnetic field, ${\bm H} =(0,0,H)$, is given by
\begin{eqnarray}
&&F[H] \nonumber \\
&&=T \sum_{\omega_q, C= \pm 1} {\rm Tr} {\rm ln} (-\hat{\tilde{L}}_{C}^{-1}(\omega_q; H)) \nonumber \\
&&=2T\sum_{\omega_q}{\rm Tr} {\rm ln} (\hat{1}-ge^{-2{\rm i} \hat{\Phi}}\hat{\Pi}(\omega_q;H)) \nonumber \\
&& + T \sum_{\omega_q}  {\rm Tr} {\rm ln} \left[\hat{1}-\left( \frac{5eHg}{4k_F^2} \right)^2 \frac{e^{-2{\rm i} \hat{\Phi}}\hat{\Pi'}^2(\omega_q ; H)}{(1-ge^{-2{\rm i} \hat{\Phi}}\hat{\Pi}(\omega_q ; H))^2}\right] . \nonumber \\
\end{eqnarray}
Then, the magnetic susceptibility is given as
\begin{eqnarray}
\chi=\chi_{dia}+\chi_{chiral},
\end{eqnarray}
where
\begin{eqnarray}
&&\chi_{dia} \nonumber \\
&&= - \left.   \frac{\partial^2}{\partial H^2} \left\{2T \sum_{\omega_q}{\rm Tr} {\rm ln} (\hat{1}-ge^{-2{\rm i} \hat{\Phi}}\hat{\Pi}(\omega_q;H)) \right\} \right|_{H \to 0} \nonumber \\
\end{eqnarray}
is the fluctuation diamagnetism term which appears also in non-chiral superconductors \cite{Galitski2001},
where the factor 2 reflects the fact that the number of fluctuation channel is two,
and 
\begin{eqnarray}
&&\chi_{chiral} \nonumber \\
&& = - \left. \frac{\partial^2}{\partial H^2}  \right.  \nonumber \\
&&  \left.  \left\{T \sum_{\omega_q} {\rm Tr} {\rm ln} \left[\hat{1}- \frac{\left( \frac{5eHg}{4k_F^2} \right)^2  e^{-2{\rm i} \hat{\Phi}}\hat{\Pi'}^2(\omega_q;H)}{\left(\hat{1}-ge^{-2{\rm i} \hat{\Phi}}\hat{\Pi}(\omega_q;H) \right)^2}\right] \right\} \right|_{H \to 0} \nonumber\\
&& = 2T \left( \frac{5eg}{4k_F^2} \right)^2  \sum_{\omega_q} {\rm Tr}  \left[\frac{ \hat{\Pi'}^2(\omega_q;H=0)}{\left( \hat{1}-g\hat{\Pi}(\omega_q;H=0) \right)^2}\right] \nonumber \\
\label{chichi}
\end{eqnarray}
is the contribution unique to chiral superconductors.
Now, we evaluate the chirality-induced  term (\ref{chichi}).
Neglecting the $\omega_q \neq 0$ terms, which are less singular in the vicinity of $T_c$ than the $\omega_q = 0$ one,
we obtain
\begin{eqnarray}
\chi_{chiral} = \frac{25 e^2T}{64 \pi k_F^4 \xi^3 (N(0) g)^2 \varepsilon^{1/2}} >0,
\end{eqnarray}
which is positive, indicating the paramagnetic response due to MC-coupling.
In this calculation, we used the momentum representation of the BPS, (\ref{L_momentum}).

\subsection{Dependence of the Results on Specific Form of $W$, Dimensionality, and Pairing Symmetry}
\label{From_of_W}
In this section, we discuss to what extent the main results, (4) and (5), depend on the functional form of
the remonetized four-point vertex, $W({\bm k}, \omega_k)$,
the spatial dimensionality, and the pairing symmetry of chiral superconducting states.

Here, we summarize the main results of this section:
1) The result that $\alpha^{Kubo}_{xy \, chiral}$ and $\sigma_{xy \, chiral}$ are proportional to $\tau^2$ is not changed 
by these three conditions.
2) The critical behavior may be changed by the dimensionality and the pairing symmetry, but not by the functional form of $W({\bm k}, \omega_k)$.
3) The magnitudes of $\alpha^{Kubo}_{xy \, chiral}$ and $\sigma_{xy \, chiral}$ depend on the specific form of $W({\bm k}, \omega_k)$:
they are decreased as the momentum-dependence is stronger.

In the succeeding subsections, we will discuss the details.

\subsubsection{{\rm $\tau$-Dependence}}

Irrespective of functional forms of $W({\bm k}, \omega_k)$,  spatial dimensionality, and the pairing symmetry of chiral superconducting states,
$\bar{A}_{C}^{i)}(\bm{ q}, \omega_q; \omega_l)$ and $\bar{S}_{C}^{i)}(\bm{ q}, \omega_q; \omega_l)$($i=a,b$) always have terms with the factor $({\rm i} \omega_l + 1/ \tau )^{-2}$
(and $({\rm i} \omega_l + 1/ \tau )^{-1}$ when $i=c$), as shown in Sec.\ref{auxiliary_explanations}.
Moreover, in any case of chiral fluctuations, the expression for the chirality-dependent fluctuation propagator,
(\ref{chiraliydepL}), holds besides numerical factors.
Therefore, it is general that $\alpha^{Kubo}_{xy \, chiral}$ and $\sigma_{xy \, chiral}$ are proportional to $\tau^2$ in clean limit.
However, their magnitudes depend on the functional form of $W$ as discussed in Sec. \ref{rangeofinteraction}

\subsubsection{{\rm Critical Behavior}}
We discuss how the critical exponents of $\alpha^{Kubo}_{xy \, chiral}$ and $\sigma_{xy \, chiral}$ as functions of
$\varepsilon=\log T/T_c$
depend on the functional form of $W({\bm k}, \omega_k)$, the dimensionality, and the pairing symmetry of chiral superconducting states.

As mentioned above, we can use the expression of
the chirality-dependent fluctuation propagator,
(\ref{chiraliydepL}) generally except numerical factors.
Then, from Eq. (\ref{alphaKuboLalpha}), we obtain
\begin{eqnarray}
\alpha^{Kubo}_{xy \, chiral} \,  \, {\rm or} \,  \, \sigma_{xy \, chiral} \propto \int d^{d} {\bm q} \frac{q^\Delta}{(\xi^2 q^2 + \varepsilon)^2} 
\label{alpha_sigma_propto_int_q_delta}
\end{eqnarray}
in the vicinity of the $T_c$,
where $\Delta$ is defined by $\bar{A}_C ({\bm q},\omega_q = 0;\omega_l) \, {\rm or} \, \bar{S}_C ({\bm q},\omega_q = 0;\omega_l)= \mathcal{O}(q^{\Delta}) $, as $q \to 0$.
Therefore we find that the spatial dimensionality and $\Delta$ determine the critical exponent.

As will be shown in Sec \ref{auxiliary_explanations},
$\Delta$ does not depend
on specific forms of $W({\bm k},\omega_k)$, but is affected by
the spatial dimensionality and pairing symmetries of chiral superconducting states.
Therefore, the critical exponents of (4) and (5) are 
independent of the specific form of $W({\bm k},\omega_k)$, though the dimensionality may change them.

In TABLE \ref{Deltadim},
we present the exponents for some typical examples.
In the case of three dimensional chiral $d_{zx} \pm {\rm i} d_{zy}$ superconducting fluctuation,
$\Delta = 2$, (see Eqs (\ref{Aa_til}-\ref{Sb_til})), and then the critical behavior is, $\alpha^{Kubo}_{xy \, chiral} \propto const - \sqrt{\varepsilon}$ (see (4)).
Moreover,
in the case of two-dimensional chiral-$p$ one
(the $d$-vector is given by ${\bm d}({\bm k})= (0,0, k_x \pm {\rm i} k_y)$, and then $\phi({\bm k}) \propto (k_x \pm{\rm i} k_y) $),
which is believed to be the paring symmetry of Sr$_2$RuO$_4$ \cite{Mackenzie},
we find, from straightforward calculations, $\bar{A}_C, \bar{S}_C \propto q^0$ and then $\alpha^{Kubo}_{xy \, chiral}, \sigma_{xy \, chiral} \propto 1/ \varepsilon$,
where the critical behavior of the former (latter) is the same as (less singular than) that of the conventional AL term in two spatial dimensions \cite{Larkin}.

\begin{table}[htb]
\begin{tabular}{l|ccc}
&\multicolumn{2}{c}{$\Delta$} \\
&0& 2& \\ \hline 
2D& $1/\varepsilon$ & $\log \varepsilon$  \\
&{\footnotesize (Sr$_2$RuO$_4$)}& \\
&&{\tiny \phantom{*}}& \\
3D & $1/\sqrt{\varepsilon}$& $const - \sqrt{\varepsilon}$  \\
&&{\footnotesize (URu$_2$Si$_2$)} 
\end{tabular}
\caption{Critical behaviors of $\alpha^{Kubo}_{xy \, chiral} $ and $\sigma_{xy \, chiral}$,
which is decided by the dimensionality and the pairing symmetry of chiral superconducting states. }
\label{Deltadim}
\end{table}

\subsubsection{{\rm Specific Form of $W$}}
\label{rangeofinteraction}

In this subsection, we discuss how the magnitudes of $\alpha^{Kubo}_{xy \, chiral}$ and $\sigma_{xy \, chiral}$ 
are influenced by the momentum-dependence of $W$.

\begin{figure}[ht]
\begin{center}
\includegraphics[width=35mm]{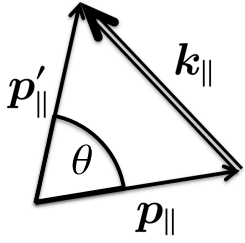} 
\caption{Relation between the relative momentums of fluctuating Cooper pairs, ${\bm p}$, ${\bm p}'$, the momentum of four-point vertex, ${\bm k}$,
and the angle $\theta$ that is the angle between ${\bm p}_{\parallel}$ and ${\bm p}'_{\parallel}$.
Here, ${\bm p}_{\parallel}$ represents the components parallel to $ab$-plane.
 }          
\label{relation_pair_cor}
\end{center}
\end{figure}

Here we consider the case of arbitrary chiral paring symmetry that is given by
$\phi ({\bm p}) \propto Y^{ m}_l(\hat{p})$ (three dimension) or $ \propto e^{  {\rm i} m \theta_{{\bm p}}}$ (two dimension),
where $Y^m_l$ is the spherical harmonic function and $\theta_{{\bm p}}$ is the angle
defined by $\tan \theta_{{\bm p}} = p_x/ p_y$.
As discussed in Sec. \ref{Derivations_of_Nernst_and_Hall},
only the odd parts of $\bar{A}_C$ or $\bar{S}_C$ with respect to time reversal, $C \to -C$, give nonzero contributions.
The integrands of $\bar{A}^{i)}_{C=1} -\bar{A}^{i)}_{C=-1} $ and $\bar{S}^{i)}_{C=1} -\bar{S}^{i)}_{C=-1}$ generally
($i=a,b,c$) 
contain the factor (see FIG. \ref{new_with_wave_number}): 
\begin{eqnarray}
&&\phi^{\dag} ({\bm p}) \phi({\bm p}') W({\bm p}-{\bm p}', \omega_k) - (\phi \leftrightarrow  \phi ') \nonumber \\
&& \propto \sin (m \theta)W({\bm p}-{\bm p}', \omega_k),
\end{eqnarray}
where $\theta$ is the angle between ${\bm p}_{\parallel}$ and ${\bm p}'_{\parallel}$.
Here, ${\bm p}_{\parallel}$ represents the projection of the vector ${\bm p}$ onto the $ab$-plane.
The relation between these vectors and angle is drawn in FIG. \ref{relation_pair_cor}.
From this equation, we find that, due to the factor $\sin (m \theta)$,
the magnitudes of $\alpha^{Kubo}_{xy \, chiral}$ and $\sigma_{xy \, chiral}$ are large
in the case that the magnitude of $W({\bm p}-{\bm p}', \omega_k)$ is large for  $\theta\sim ({\rm half \,  odd \,  integer} ) \times (\pi/m)$.

We, now, consider a simple case that
the momentum dependence of $W({\bm k}, \omega_k)$ 
has a dominant peak at ${\bm k} = {\bm Q}_0$ with width $1/\xi_{{\bm Q}_0}$.
When the peak is sharp, 
the domain of integration that contributes to $\alpha^{Kubo}_{xy \, chiral}$ and $\sigma_{xy \, chiral}$ 
is restricted to the region in which ${\bm p}-{\bm p}' \sim {\bm Q}_0$.
However, it is quite exceptional
that the angle $\theta$ is nearly equal to $ ({\rm half \, odd \,  integer} ) \times (\pi/m)$
when ${\bm p}$ and ${\bm p}'$ satisfy the above condition.
Therefore, generally, the magnitudes of $\alpha^{Kubo}_{xy \, chiral}$ and $\sigma_{xy \, chiral}$
are small when $\xi_{{\bm Q}_0}$ is large.
On the other hand, when $\xi_{{\bm Q}_0}$ is small, 
the domain of integration in which $\theta \sim ({\rm half \, odd \,  integer} ) \times (\pi/m)$,
is included in the domain in which $W({\bm p}-{\bm p}', \omega_k)$ has large values,
and then the magnitudes of $\alpha^{Kubo}_{xy \, chiral}$ and $\sigma_{xy \, chiral}$ are larger than the case of large 
$\xi_{{\bm Q}_0}$.
As a result,
the magnitudes of $\alpha^{Kubo}_{xy \, chiral}$ and $\sigma_{xy \, chiral}$ increase as $\xi_{{\bm Q}_0}$ become smaller.

Finally, we consider the case of URu$_2$Si$_2$.
In this case, the pairing function is given by $\phi ({\bm p}) \propto Y^{ 1}_2(\hat{p})$ and
the interaction is mediated via short-range antiferromagnetic spin-fluctuation with ${\bm Q}_0 = (0,0,2\pi/a^{lattice}_z)$, where $a^{lattice}_z$ is the lattice constant.
Here $\xi_{{\bm Q}_0}$ is the correlation length of this fluctuation.
Since 
$\theta$ is zero when ${\bm Q}_0 = {\bm p}-{\bm p}'$, then,
the magnitudes of $\alpha^{Kubo}_{xy \, chiral}$ and $\sigma_{xy \, chiral}$ become small,
if this fluctuation were to be long-range.

\subsubsection{{\rm Auxiliary Explanations}}
\label{auxiliary_explanations}

\paragraph{\rm Factors $({\rm i} \omega_l + 1/ \tau )^{-2}$ and $({\rm i} \omega_l + 1/ \tau )^{-1}$ in $\bar{A}_{C}^{i)}(\bm{ q}, \omega_q; \omega_l)$ and $\bar{S}_{C}^{i)}(\bm{ q}, \omega_q; \omega_l)$}
\indent

Here, we show that in any case of the functional form of $W$, the dimensionality, and the pairing symmetry of chiral superconducting states,
$\bar{A}_{C}^{i)}(\bm{ q}, \omega_q; \omega_l)$ and $\bar{S}_{C}^{i)}(\bm{ q}, \omega_q; \omega_l)$ for $i=a$, $b$ always have the terms proportional to $({\rm i} \omega_l + 1/ \tau )^{-2}$
(and $({\rm i} \omega_l + 1/ \tau )^{-1}$ for $i=c$).

\begin{figure}
\begin{center}
 \includegraphics[clip, width=35mm]{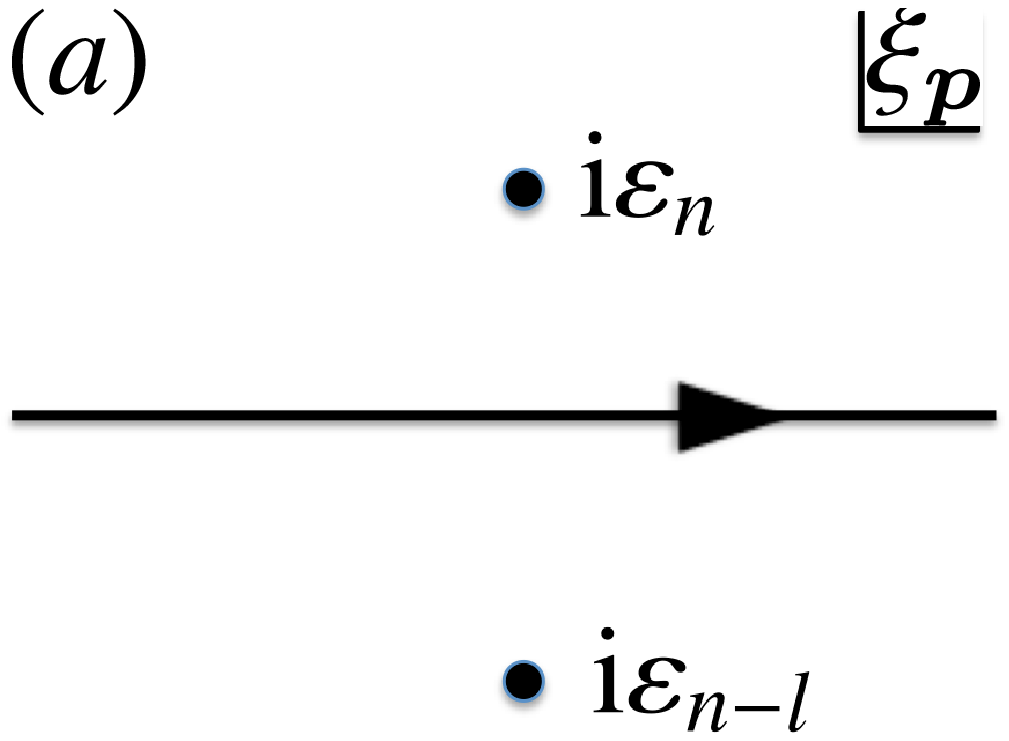}\\
 \includegraphics[clip, width=80mm]{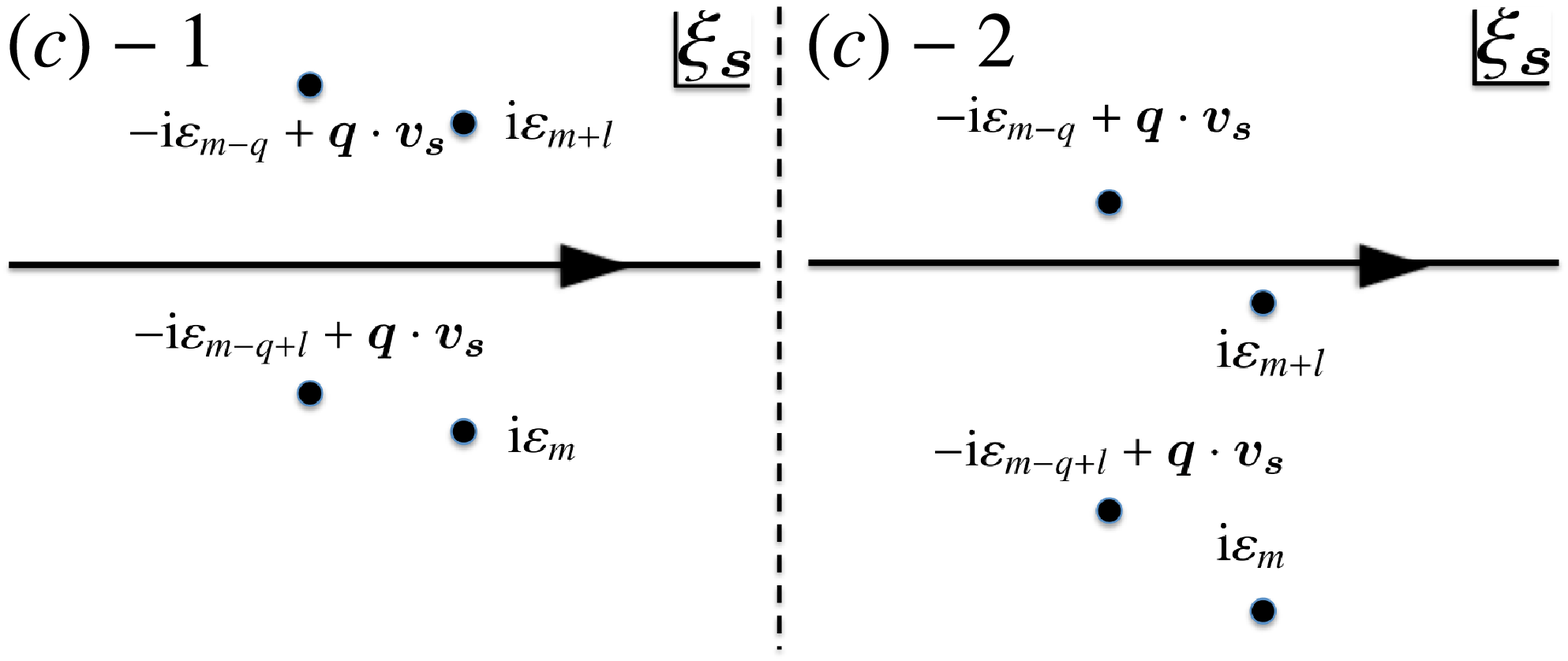}
 \centering
 \caption{The contours of energy-integration in Eqs. (\ref{A_Ca_S_Ca_general}) (upper panel) and (\ref{A_Cc_S_Cc_general}) (lower panels). 
 Dots represent poles of order 1.
 Each contour integration generates the factor $({\rm i} \omega_l + 1/ \tau )^{-1}$.}          
 \label{contours}
 \end{center}
 \end{figure}

First, for $i=a$, in any case, we can write (see FIG. \ref{new_with_wave_number} ($a$)):
\begin{eqnarray}
&& \bar{A}^{a)}_C({\bm q}, \omega_q; \omega_l) \, \, {\rm or} \, \, \bar{S}^{a)}_C({\bm q}, \omega_q; \omega_l) \nonumber \\
&& = \sum_{{\bm p}, {\bm s},{\bm q},n,m} G({\bm p}, \varepsilon_{n-l}) G({\bm p}, \varepsilon_n) G({\bm s}, \varepsilon_{m-l}) G({\bm s}, \varepsilon_m) \nonumber \\
&& \qquad \qquad \times g^a ({\bm p},{\bm s}, n , m;l), \label{A_Ca_S_Ca_general}
\end{eqnarray}
where $g^a$ is a certain function.
Then, integrating it over the energy $\xi_{{\bm p}}$ along the contour shown in FIG. \ref{contours} ($a$), we obtain the factor
$({\rm i} \varepsilon_{n}- {\rm i} \varepsilon_{n-l})^{-1}= ({\rm i} \omega_l + 1/ \tau )^{-1}$ at $n=0,1,\cdots l-1$,
and this factor does not appear for other values of $m$.
Similarly, the integration over $\xi_{{\bm s}}$ also generates the factor $({\rm i} \omega_l + 1/ \tau )^{-1}$ at $m=0,1,\cdots l-1$.
Therefore, $\bar{A}_{C}^{a)}(\bm{ q}, \omega_q; \omega_l)$ and $\bar{S}_{C}^{a)}(\bm{ q}, \omega_q; \omega_l)$
have terms with factor $({\rm i} \omega_l + 1/ \tau )^{-2}$.
We can prove this relation also for the case of $i=b$ by using a similar argument.

Next,
for $i=c$, we can write (see FIG. \ref{new_with_wave_number} $(c)$-1):
\begin{eqnarray}
&& \bar{A}^{c)}_C({\bm q}, \omega_q; \omega_l) \, \, {\rm or} \, \, \bar{S}^{c)}_C({\bm q}, \omega_q; \omega_l) \nonumber \\
&& = \sum_{ {\bm s},{\bm p},m,n} G({\bm s}, \varepsilon_{m+l}) G({\bm s}, \varepsilon_m)  \nonumber \\
&& \qquad \quad \times  G({\bm q} - {\bm s}, -\varepsilon_{m-q}) G({\bm q} - {\bm s}, - \varepsilon_{m-q+l})  \nonumber \\
&& \qquad \quad \times  G({\bm q} - {\bm p}, -\varepsilon_{n-q+l}) G({\bm p}, \varepsilon_{n+l}) \nonumber \\
&& \qquad  \quad \times g^c({\bm s}, {\bm p},{\bm q}, m;l), \label{A_Cc_S_Cc_general}
\end{eqnarray}
where $g^c$ is a certain function.
When $\max \{ -l , q-l \} \leq m \leq \min \{ -1 , q-1 \}$,
the pole structure of the complex $\xi_{{\bm s}}$-plane is given by FIG. \ref{contours} ($c$),
and then the integrating over $\xi_{{\bm s}}$ generates the terms proportional to $({\rm i} \omega_l + 1/ \tau )^{-1}$.
Such terms appear also for the value of $m$, at which three poles exist in the upper-half plane or lower-half-plane and
the other one exits in the other side (FIG. \ref{contours} ($c$)-2),
and do not appear when all poles exist on the same side.
On the other hand, integrating over $\xi_{{\bm p}}$ generates no term proportional to $({\rm i} \omega_l + 1/ \tau )^{-1}$.
Therefore, $\bar{A}_{C}^{c)}(\bm{ q}, \omega_q; \omega_l)$ and $\bar{S}_{C}^{c)}(\bm{ q}, \omega_q; \omega_l)$
have terms with a factor $({\rm i} \omega_l + 1/ \tau )^{-1}$.

 \phantom{*}

\paragraph{\rm Critical exponent $\Delta$ independent of specific form of $W$}
\indent

In this subsection, we show that the critical exponent $\Delta$ is independent of the specific form of $W({\bm k}, \omega_k)$.
We consider the case that the interaction is repulsive, i.e. $W({\bm k},\omega_k)>0$.

For every $i=a,b,c$,
we can write $\bar{A}^{i)}_C$ or $\bar{S}^{i)}_C$ as
\begin{eqnarray}
&&\bar{A}^{i)}_{C}(\bm{ q}, \omega_{q}=0; \omega_l)  \, \, {\rm or} \, \, \bar{S}^{i)}_{C}(\bm{ q}, \omega_{q}=0; \omega_l) \nonumber \\
&&= \sum_{{\bm k}, \omega_k} h({\bm q},{\bm k}, \omega_k ; \omega_l) W({\bm k}, \omega_k).
\end{eqnarray}
The important point is that $W$ is independent of ${\bm q}$. Then,
expanding $h({\bm q},{\bm k}, \omega_k ; \omega_l) $ with respect to ${\bm q}$, we obtain
\begin{eqnarray}
&& \bar{A}^{i)}_{C}(\bm{ q}, \omega_{q}=0; \omega_l) \, \, {\rm or} \, \, \bar{S}^{i)}_{C}(\bm{ q}, \omega_{q}=0; \omega_l)  \nonumber\\
&& =  \sum_{{\bm k}, \omega_k} h(\bm{k}, \omega_{k}; \omega_l) W({\bm k}, \omega_k) \nonumber \\
&& + q^{2}_{x} \sum_{{\bm k}, \omega_k} h_{xx}(\bm{k}, \omega_{k}; \omega_l) W({\bm k}, \omega_k) \nonumber \\
&& + q_{y}^{2} \sum_{{\bm k}, \omega_k} h_{yy}(\bm{k}, \omega_{k}; \omega_l) W({\bm k}, \omega_k) \nonumber \\
&& + \cdots.
\end{eqnarray}
Here, we neglect terms that vanish after performing the integration, (\ref{alpha_sigma_propto_int_q_delta}).
$\Delta$ is the lowest number of dimensions with respect to ${\bm q}$ of the term in which $\sum h_* W$ is nonzero,
for $*= \, \, , xx,yy,zz,xxxx,xxyy, \cdots$.
Since $W$ is positive definite, $\sum h_* W$ leads to $\sum  h_* W'=0$, where $W'$ is another form of potential energy that is positive definite.
Therefore, $\Delta$ for some particular form of $W$ is the same as that of another one, $W'$,
and thus, we conclude that $\Delta$ is independent of the specific form of $W$.

\end{document}